\documentclass[prl,reprint,preprintnumbers,amsmath,amssymb]{revtex4-2}
\usepackage{graphicx}
\usepackage{dcolumn}
\usepackage{bm}
\usepackage{hyperref}
\usepackage{physics}
\begin{document}

\title{Active Suppression of Quantum Dephasing in Resonantly Driven Ensembles}  
\author{C. He and R.R. Jones}     
\affiliation{Department of Physics, University of Virginia, Charlottesville, VA 22904-4714} 

\date{\today}

\begin{abstract}
We have used quantum control to suppress the impact of random atom positions on coherent population transfer within atom pairs, enabling the observation of dipole-dipole driven Rabi oscillations in a Rydberg gas with hundreds of atoms. The method exploits the reduced coupling-strength sensitivity of the off-resonant Rabi frequency, and coherently amplifies the achievable population transfer in analogy to quasi-phase-matching in non-linear optics. Simulations reproduce the experimental results and demonstrate the potential benefits of the technique to other many-body quantum control applications.
\end{abstract}

\maketitle

The ability to create and manipulate superposition states possessing well-defined relative amplitudes and phases is a key capability for controlling quantum systems. Typically, (near) resonant interactions play an essential role, enabling the transfer of probability amplitude between states to establish an initial coherence or entanglement, manipulate it, and/or measure it in a desired quantum operation. The application of a coherent coupling over a specific period of time creates a superposition in which the relative amplitude and phase of the constituent states is fully determined by the interaction strength and detuning from resonance. Thus, in an ensemble of nominally identical elements for which the interaction strength or detuning is non-uniform, the characteristics of the quantum-state created (or modified) through the interaction will vary among the constituents. Alternatively, if the quantum-state of interest involves multiple elements of the ensemble, such inhomogeneities can significantly impact the evolution of entanglement and the nature of the quantum-state \cite{blkexcite,latticedicke,latticedicke2,Rydsqueeze,superrad,Rinduce}. 

We have developed a control scheme that can substantially reduce the impact of inhomogeneities on quantum ensembles driven by (near) resonant interactions. The method is applicable when, for all elements, the relevant resonance occurs at the same value of some externally tunable parameter, but where the coupling strength varies. Common examples of this scenario include optical excitation of a spatially extended sample using a laser beam with a non-uniform spatial intensity distribution \cite{Adiapassage,CohExt,CohExtR}, or F{\"o}rster-resonant interactions \cite{Forster,ForsterR,2atomForster} between non-uniformly spaced Rydberg atoms in a frozen gas \cite{Anderson,Mourachko,Robicheaux,Westerman,Raithel,Ryabtsev,cusp}. We focus on the latter. 

\begin{figure}
\centering
\resizebox{0.45 \textwidth}{!}{\includegraphics{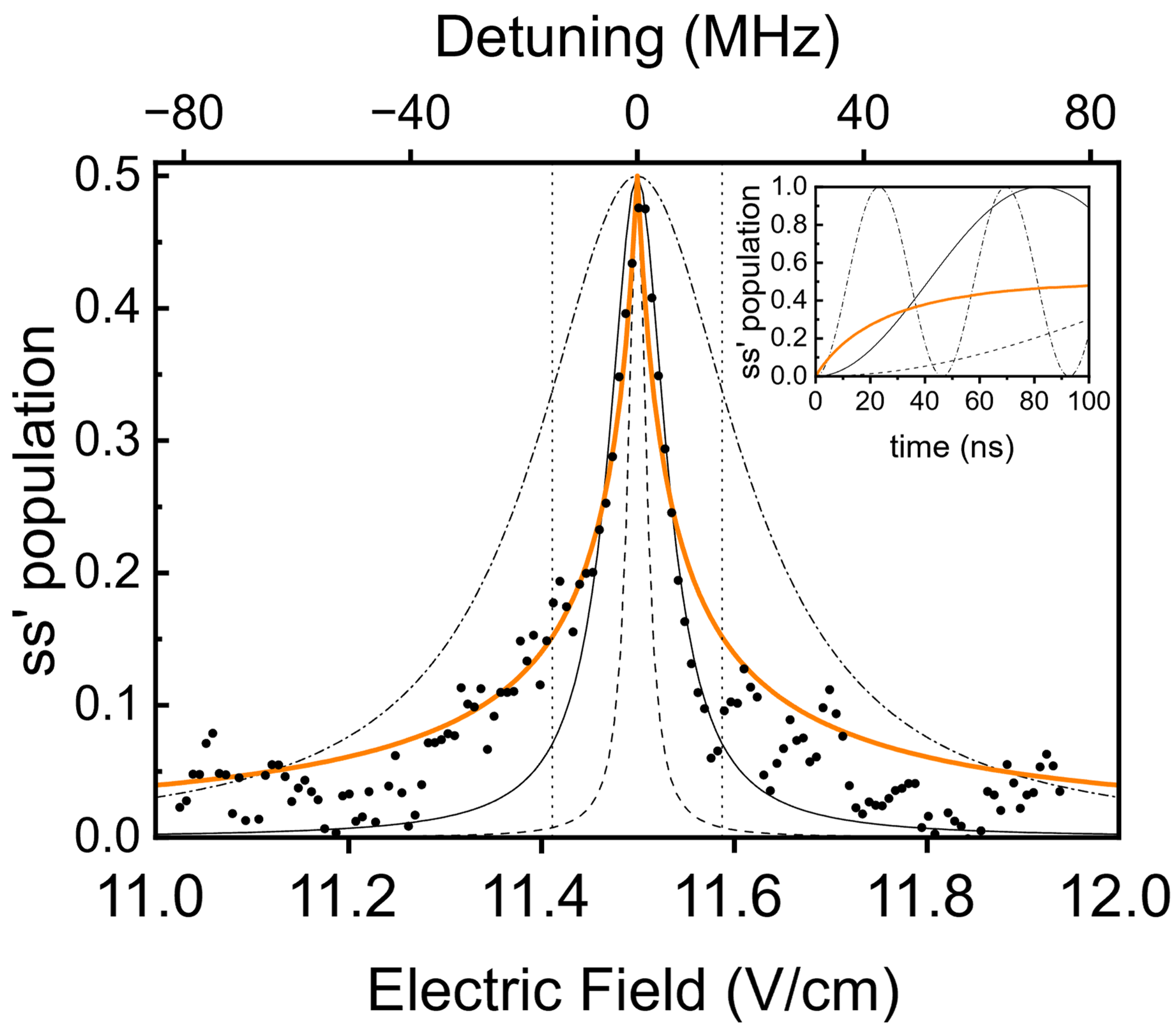}}
\caption{Dipole-dipole driven population transfer probability vs. applied electric field (lower axis) and bare-state energy splitting (upper axis). (Black dots) Experimental result for a 500ns interaction time and $\rho$= 2 $\times$ 10$^{9}$ cm$^{-3}$. (Solid orange) Simulated cusp lineshape for the ensemble \cite{cusp}, including nearest-neighbor Rydberg interactions only \cite{SM}. (Thin black lines) Predicted Lorentzian lineshapes for atom pairs with DD couplings corresponding to 20$\%$ (dashed), 50$\%$ (solid), and 80$\%$ (dot-dashed) levels in the ensemble integrated coupling-strength probability distribution. The vertical dotted lines mark the $\pm$ detuning points for a typical control sequence. Inset: Simulated population transfer vs interaction time with the system tuned to resonance for 500ns, and line types corresponding to those in the main figure.  }
\label{fig1}
\end{figure}

Specifically, we have used coherent control of near-resonant dipole-dipole (DD) interactions in a cold, random many-atom Rydberg gas to actively suppress dephasing associated with the variation in the coupling strength between neighboring atoms. To implement the control, pulsed electric field sequences rapidly tune the eigenstates of Rb Rydberg atom pairs back and forth across a F{\"o}rster-resonance, holding the atoms in opposing wings of the resonance for equal times, substantially reducing the variation in the generalized Rabi frequency over the ensemble. The sequence results in a periodic reversal of the sign of the Rabi phase-lag accrued in adjacent time intervals, amplifying the coherent population transfer in a time-domain analogy to spatial quasi-phase matching schemes in non-linear materials \cite{QPM1}, and enabling what is to our knowledge the first observation of {\it DD-driven} Rabi oscillations in a random Rydberg gas with more than a few atoms \cite{2atomForster}. The mechanism underlying the control can be understood using an analytic two atom model, but numerical simulations that include beyond nearest neighbor interactions are required to obtain quantitative agreement with the experiments. Simulations also demonstrate the effectiveness of the technique for suppressing variations in the Rabi phase and, accordingly, in the quantum state distribution in spatially ordered Rydberg arrays with small residual differences in atom separation, such as those being employed for quantum simulation and computing applications \cite{QC1,QC2,QC3}.

In the experiments, a random ensemble of $\sim$ 1000 32$p_{3/2}, |m_j|$=3/2 $^{85}$Rb Rydberg atoms is excited from a 70 $\mu$K magneto optical trap (MOT) using a 300ns laser pulse \cite{SM}. Voltages applied to a pair of parallel plates straddling the MOT facilitate the creation of an initial static, $F \simeq $ 12 V/cm, and subsequent pulsed, electric fields within the excitation volume. Following the Rydberg excitation, electric field steps with fast ($\sim$ 2 ns) rise and fall times rapidly Stark tune the Rydberg atoms on, or about, the 32$p_{3/2},|m_j|$=3/2$\,+\,$32$p_{3/2}, |m_j|$=3/2 $\leftrightarrow$ 32$s\,+\,$33$s$ (i.e. $pp\leftrightarrow$ $ss'$) F{\"o}rster-resonance near $F_0 =$11.5V/cm \cite{Westerman}. Time spent in the vicinity of the resonance enables coherent population transfer between $pp$ and $ss'$ atom pairs via an anisotropic DD interaction $V(R,\Theta)=r_{ps}r_{ps'} f(\Theta)/R^{3}$, where $r_{ps}$ ($r_{ps'}$) are radial matrix elements between 32$p$ and 32$s$ (33$s$), $R$ is the separation between the atoms in each pair, and $f(\Theta)$ describes the anisotropy \cite{Noel} in terms of the angle $\Theta$ between the internuclear and electric field axes \cite{SM}. After the atoms have interacted for the desired time, state-selective field ionization is used to measure the final population in different Rydberg states \cite{tfg}. 

Figure \ref{fig1} shows the $pp$ to $ss'$ population transfer probability (i.e. the $ss'$ population normalized to the total Rydberg population) as a function of the tuning field, for a 500ns interaction time and Rydberg density $\rho=2\times10^{9}$ cm$^{-3}$. The agreement between the simulated ensemble and measured lineshapes is reasonable. The simulation and experiment differ more in the lineshape wings, where the transition probability is dominated by atom pairs with separations much less than average \cite{cusp}. While the agreement there can be improved somewhat by modeling the effect of Rydberg blockade \cite{blockade}, which suppresses the excitation of the closest atom pairs, our principal measurements focus on interactions closer to resonance where blockade effects have negligible effect \cite{SM}. The half-width at half-maximum (HWHM) energy width of the ensemble cusp and 50\% Lorentzian lineshapes are identical, $E_0 = 2 V_0$ ($\simeq$ 3 MHz for $\rho$ = 1$\times$10$^9$cm$^3$), where $V_0$ is the angle averaged interaction strength at the most probable pair separation for a given density, $R_0\simeq(2\pi\rho)^{-1/3}$.

\begin{figure}
\centering
\resizebox{0.35 \textwidth}{!}{\includegraphics{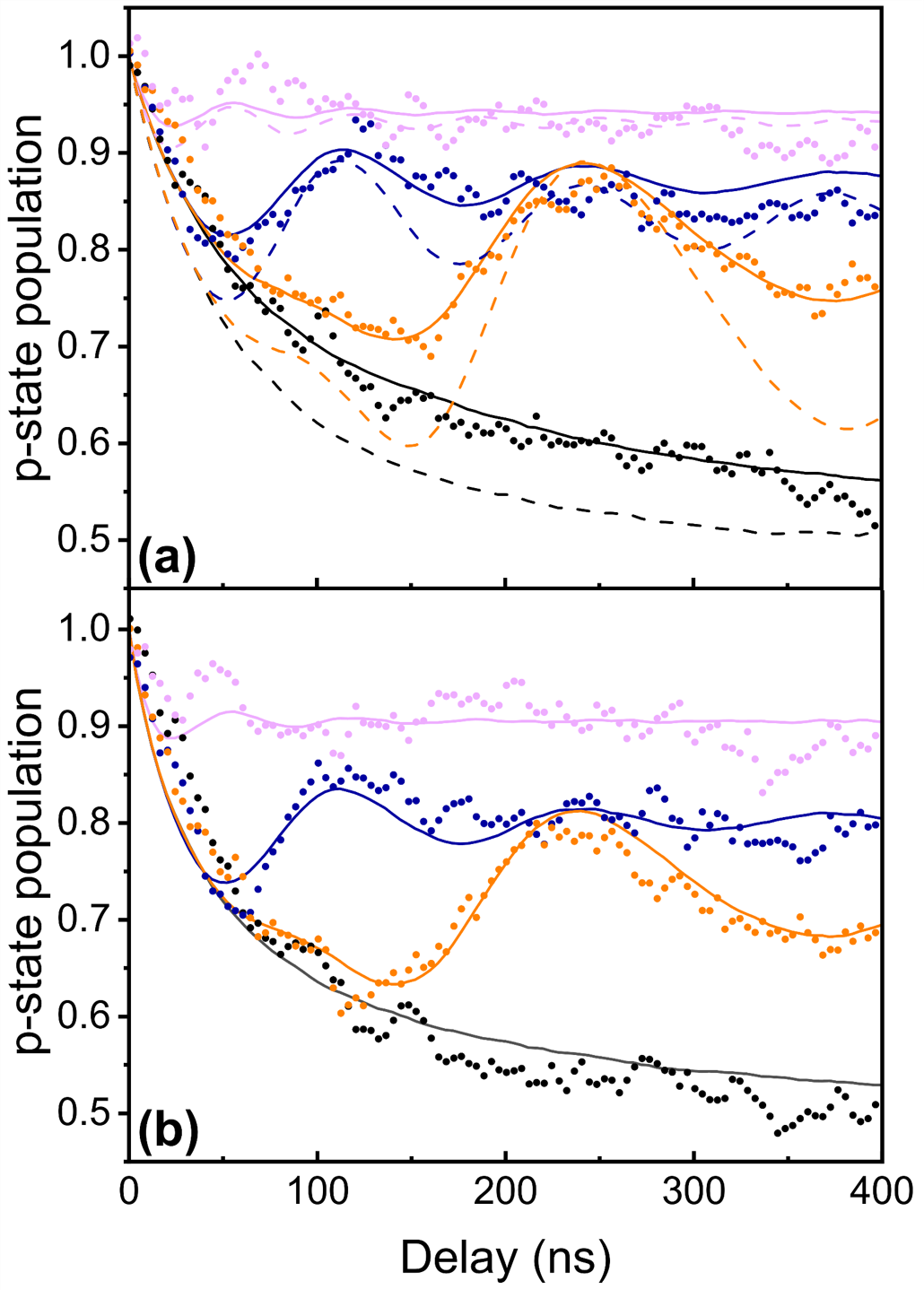}}
\caption{Experimental results (dots), along with 2-atom (dashed lines, Fig. 2a only) and 4-atom (solid lines) simulations, for $p$ state population (normalized to total population) as a function of delay $T$ after an ensemble of initially excited $p$ atoms with peak Rydberg density (a) 1$\times10^9$/cm$^3$ and (b) 2$\times10^9$/cm$^3$ is tuned on-resonance (black) or $+$15MHz (magenta) on the positive field side of resonance for $t>0$, or subjected to QPM sequences with a total of $N$=2 (blue) or $N$=4 (orange) time zones with alternating detunings of $\pm$15MHz. In (a) and (b), $E/E_0 \simeq$ 5 and 2.5, respectively. There are no adjustable parameters in the simulations. All data in each graph have been scaled using a single normalization constant.}
\label{fig2}
\end{figure}

\begin{figure*}
\centering
\resizebox{0.32\textwidth}{!}{\includegraphics{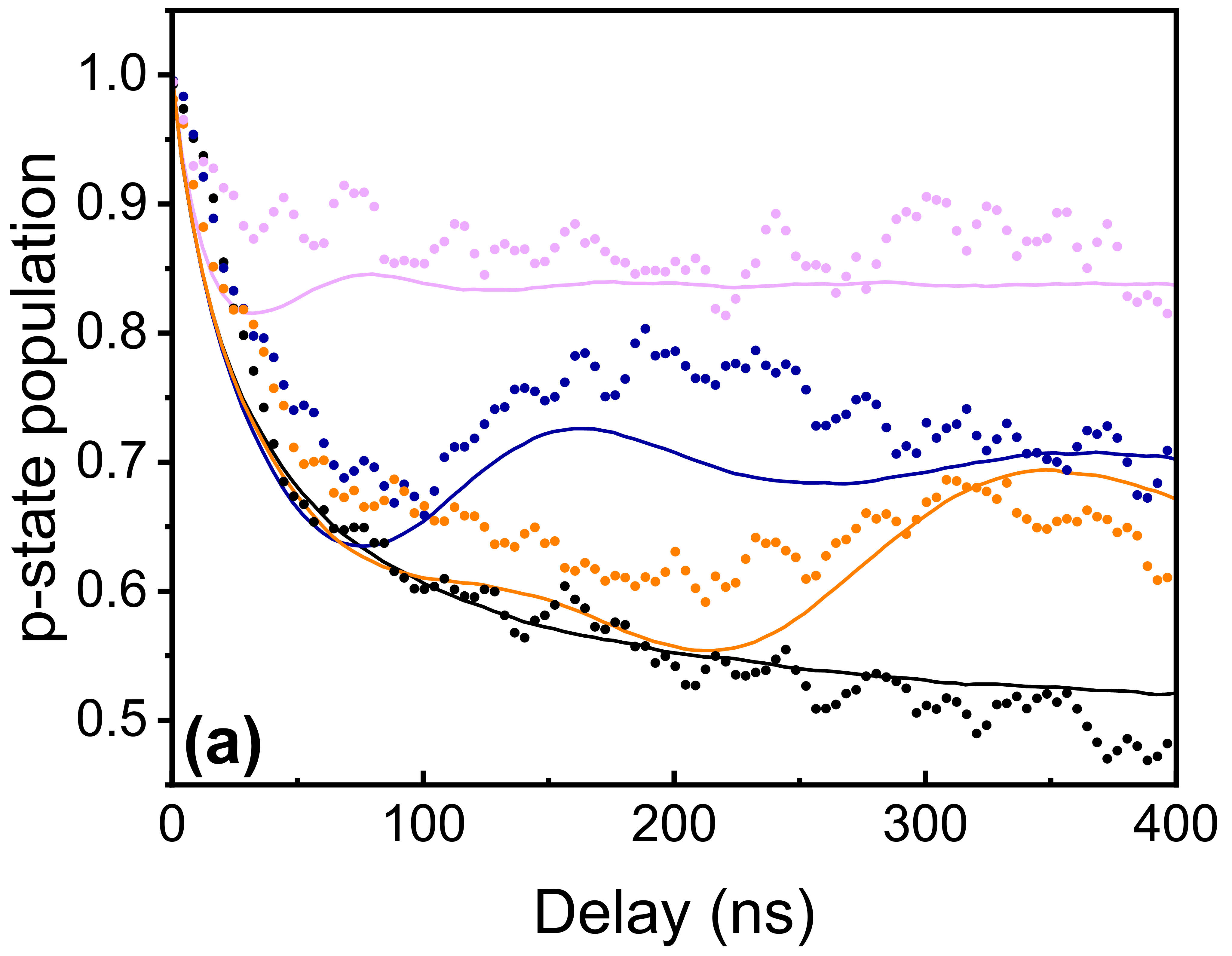}}
\resizebox{0.32\textwidth}{!}{\includegraphics{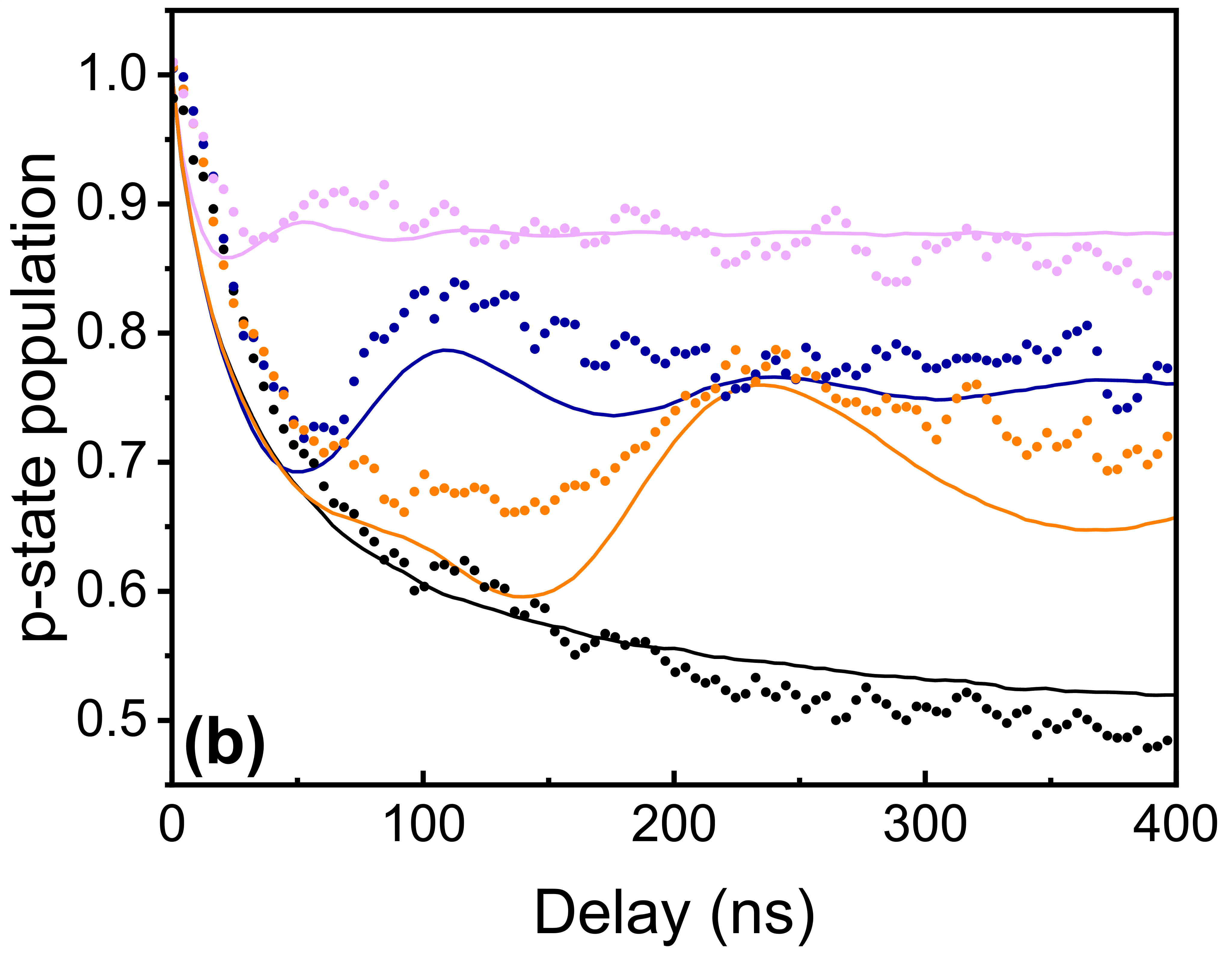}}
\resizebox{0.32\textwidth}{!}{\includegraphics{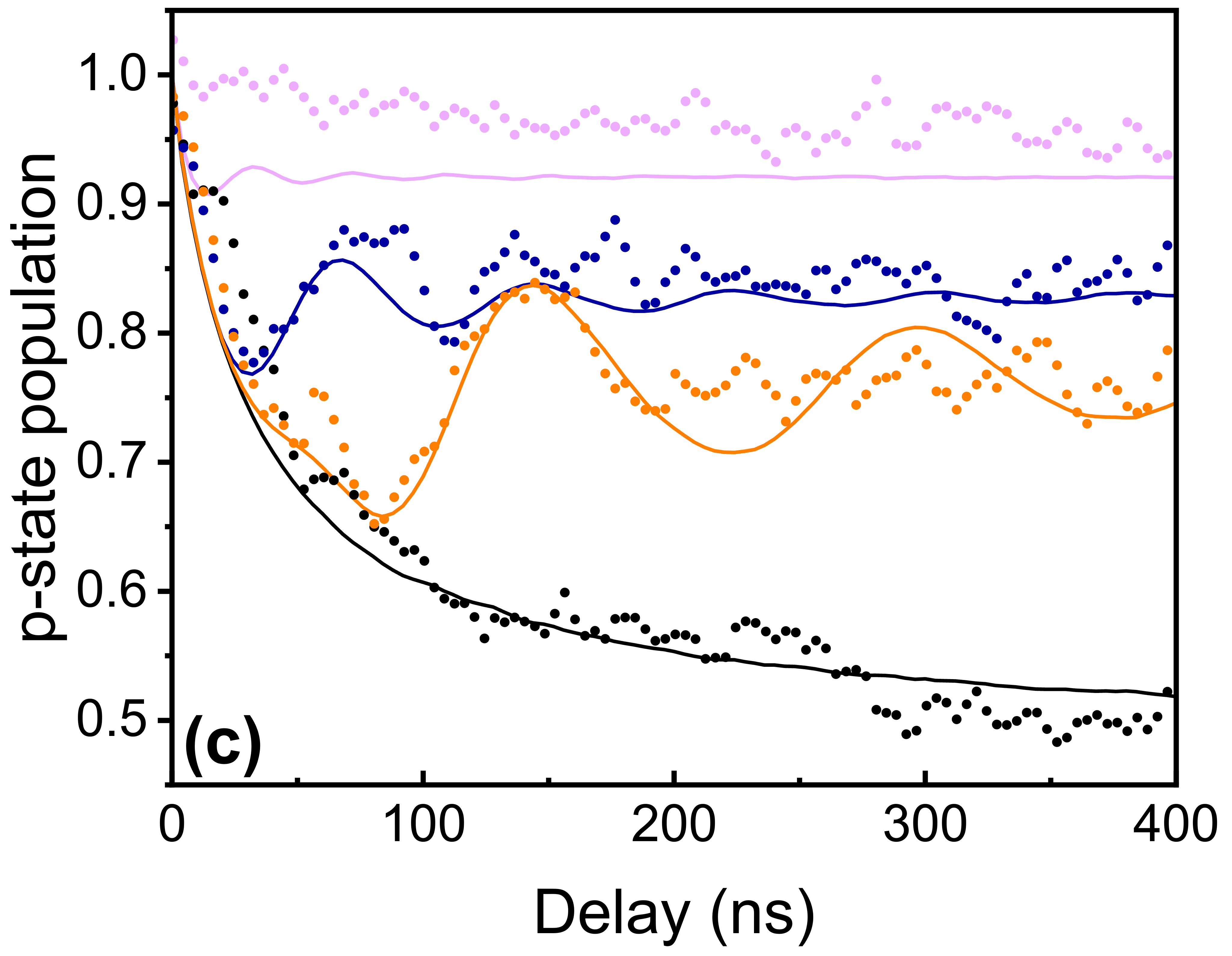}}
\caption{Analogous to Fig. \ref{fig2} except that the Rydberg density is fixed at $3\times10^9$/cm$^3$ ($E_0 \simeq$ 9MHz) with detuning magnitudes of (a)$E=$10MHz, (b)$E=$15MHz, and (c)$E=$25MHz.}
\label{fig3}
\end{figure*}

The inset to Figure \ref{fig1} shows the predicted evolution of the $ss'$ population when the atoms are tuned to the F{\"o}rster-resonance and allowed to interact for 500ns. The curves corresponding to isolated atom pairs exhibit clear Rabi oscillations with frequencies $\gamma(R,\Theta) = 2 V(R,\Theta)$ \cite{2atomForster,3}. Due to the variation in $\gamma(R,\Theta)$, the predicted evolution for the ensemble exhibits a monotonic increase and saturation, but no observable Rabi oscillations. Measurements of the remaining $pp$ population in an experimental ensemble show a corresponding monotonic decrease (black dots in Figs. \ref{fig2} and \ref{fig3}). 

The situation is similar when the atom pairs are detuned from resonance by an energy $E$. The oscillations in the transition probability for individual atom pairs, at generalized Rabi frequencies $\Gamma(R,\Theta)=\sqrt{E^2+4V^2}$ \cite{3}, are again obscured in the ensemble average, leading to an initial increase (decrease) and then saturation of the $ss'$ ($pp$) population (magenta dots in Figs. \ref{fig2} and \ref{fig3}). The primary differences from the on-resonant case being a decrease in the maximum ensemble-averaged transition probability, $\langle4V^2/\Gamma^2\rangle$, and a more rapid saturation due to the increase in $\Gamma$.

In the remainder of this Letter we use the visibility of Rabi oscillations as a metric for the effectiveness of control sequences for suppressing the impact of inhomogeneities in the coupling strength on coherent state preparation across the ensemble. Our principal goal is to reduce the variation $\Delta\phi$ in the Rabi phase, $\phi=\Gamma T$, acquired by different atom pairs interacting for the same time $T$, enabling the creation of (more) uniform distributions of arbitrary quantum superpositions of $pp$ and $ss'$ pair states. 

Since the variation $\Delta V$ in the interaction strength within the random ensemble is comparable to $V_0$, the half-width $E_0$ of the ensemble-averaged population transfer cusp lineshape provides a reasonable measure of $\Delta V$. On resonance, the phase variation $\Delta\phi=E_0T$ is a maximum, resulting in the greatest dephasing for a given interaction time $T$. However, if all atom pairs have the same large detuning $|E|\gg E_0$, then $\Delta\phi\approx T{E_0}^2/(2E)$, and there is negligible dephasing for times $T\ll 2E/{E_0}^2$. Of course, with large detunings, the maximum ensemble averaged transition probability, $\langle 4V_{0}^2/\Gamma^2\rangle\sim{E_0}^2/E^2$, is also negligibly small. Therefore, interactions at large detuning alone are not an option for creating ensembles of atoms in identical arbitrary quantum superposition states.

Off-resonance, the transition amplitude is limited by the relative phase-mismatch between the coupled states, which advances at a rate, $E$. A similar effect caps optical harmonic conversion in non-linear crystals where harmonic and fundamental waves travel with different indices of refraction. In that case, quasi-phase-matching (QPM) via periodic poling \cite{QPM1} characterized by a reversal of the sign of the non-linear susceptibility at regular intervals within the crystal, can dramatically increase the conversion efficiency. We demonstrate an analogous approach, periodically flipping the sign of $E$, to implement a time-domain variant of QPM in a driven quantum system. {\it This enables large transition amplitudes at any detuning.}

Figures \ref{fig2} and \ref{fig3} show the evolution of the $pp$ character when QPM sequences are applied to a DD-driven many-atom Rydberg gas. During the sequence, the initial 32$p$ ensemble is subjected to periodic electric field steps or ``jumps" that rapidly ($\sim$ 2ns) tune the atom pairs back and forth across the resonance, reversing the sign of $E$. The atoms interact with alternating detunings $\pm E$ in $N$ adjacent time ``zones", each with a duration $T/N$, for a total time $T$ \cite{1}. 

Several features of the data in Figs. \ref{fig2} and \ref{fig3} illustrate the effectiveness of the control sequences for reducing dephasing within the ensemble while allowing for large probability amplitude transfers. First, the level of off-resonant coherent population transfer is substantially enhanced through QPM, increasing with $N$ so that large detunings can be used to reduce dephasing without limiting the range of constituent state amplitudes that can be realized. Second, Rabi oscillations that are completely obscured in the on-resonant (black) and constant detuning (magenta) data are revealed through QPM. To our knowledge this is the first observation of DD-driven Rabi oscillations in a random ensemble with more than a few atoms. As expected for large detunings, the period of the damped Rabi oscillations is largely independent of density (Fig. \ref{fig2}) and inversely proportional to $\Gamma\simeq E$ (Fig. \ref{fig3}). Interestingly, the oscillation period and damping time increase in proportion to the number of QPM zones.

The principal characteristics of the data in Figs. \ref{fig2} and \ref{fig3} are captured by numerical simulations \cite{He,SM}. The calculations follow the quantum evolution of 5000 individual groups of two (dashed lines) or four (solid lines) nearest neighbor 32$p$ atoms that are selected from within a random ensemble at the measured densities. The atoms in each group are allowed to interact for a total time $T$ and the probabilities for finding atoms in the initial 32$p$ state are summed over the individual groups. All atom pairs are subject to the (near) resonant $pp\leftrightarrow ss'$ F{\"o}rster interaction for the constant or periodically reversing detuning. Pairs of atoms within the four atom groups are also subject to field-independent DD-exchange interactions, $ps \leftrightarrow sp$ and $ps' \leftrightarrow s'p$ \cite{SM}, which are partially suppressed by the resonant interactions between nearest neighbors \cite{Robicheaux,Raithel,He}. While ignoring exchange provides qualitative agreement with the data (dashed lines in Fig. \ref{fig2}a), a quantitative comparison requires its inclusion in the 4-atom model \cite{He,SM}.

The good quantitative agreement between experiment and theory in Fig. \ref{fig2} suggests that the few atom model captures the essential physics underlying the evolution of the DD-coupled Rydberg gas with, and without, the control sequences, particularly at lower Rydberg densities. At the higher density used in Fig. \ref{fig3}, the accuracy of the 4-atom model is reduced and the agreement with experiment is less impressive. Specifically, the increase in atom-atom coupling strength decreases the relevant interaction time-scales so that the 2ns duration of the field step is not totally negligible. In addition, excitation hopping outside of 4-atom groups is more probable at earlier times within the 400ns observation window. Still, the agreement is sufficient to support our interpretation of the control dynamics. Indeed, more insight into how QPM actively suppresses dephasing, the explicit form of the observed $N$-dependence of the amplitude and frequency of the Rabi oscillations, and even our rationale for describing the observed modulations as Rabi oscillations (rather than a more generic interference effect) is better obtained from approximate analytic expressions describing the 2-level quantum state evolution within isolated nearest-neighbor atom pairs.

Applying the standard state-transformation for a coherently coupled 2-level system, the population transfer probability from $pp$ to $ss'$ exhibits Rabi oscillations, $P_{ss'}(T) = (\frac{2V}{\Gamma})^2 \sin^2{(\frac{1}{2} \Gamma T)}$ as a function of the interaction time $T$ \cite{Meystre,SM}. Successive application of that state transformation, describing a QPM sequence in which the sign of the detuning alternates in $N$ successive time zones of duration $T/N$ (assuming $E\gg V$ and $P_{ss'}(T)\ll 1$), obtains an identical expression for $P_{ss'}(T)$, provided $\Gamma$ is replaced with $\Gamma/N$ \cite{SM}. Thus, we refer to the observed modulations as Rabi oscillations. The predicted proportionality between the Rabi period and $N$ is a clear feature of the data and simulations in Figs. \ref{fig2} and \ref{fig3}. The observed enhancement in $P_{ss'}(T)$ is not as large as the predicted factor of $N^2$, due to a breakdown of the assumption $P_{ss'}(T)\ll 1$ (also responsible for the deviation from purely sinusoidal modulations) and to non-negligible contributions from atom pairs with smaller than average separations (and $E\sim V$) \cite{2}.

Two principal factors are responsible for the suppression of dephasing through QPM. First, as discussed previously, the variation in $\Gamma$ is significantly smaller for $|E|\gg E_0$, resulting in a substantial reduction in the ensemble phase variation $\Delta\phi$ over any time interval. Second, since the phase evolution is reversed in successive time zones (similar to a spin echo \cite{Hahn_echo}), $\Delta\phi$ does not accrue over the total interaction time $T$. The effect is distinct from an echo, however, because the coupling is present throughout the system evolution and the relative phase $\varphi$ between the superimposed $pp$ and $ss'$ states advances at a non-constant rate within each zone. At the end of each zone in a QPM sequence, the magnitude of $\Delta\phi$ is equal to that at the completion of zone 1, acquired during the interval $T/N$ \cite{SM}. Accordingly, if the ensemble dephases after a time $T=\tau$ while at constant detuning then, with a QPM sequence, it will not dephase until the time spent in zone 1 is $T/N=\tau$, i.e. until the total interaction time is $T=N\tau$. Thus, QPM extends the dephasing time by a factor of $N$.

Figs. \ref{fig2} and \ref{fig3} clearly show the predicted extension of the dephasing time with increasing $N$. Interestingly, since the dephasing time and Rabi frequency are proportional to $N$ and $1/N$, respectively, their product which gives the number of Rabi cycles that can be observed within the dephasing time, is independent of $N$. The data and simulation show that this relationship continues to hold for $E\sim E_0$ and $P_{ss'}(T)\sim 1$.  

QPM sequences can be even more effective at suppressing dephasing in systems with narrower coupling strength distributions. One example is an ensemble of resonantly-driven atoms near the center of a Gaussian laser beam \cite{Adiapassage,CohExt,CohExtR}. Another involves interactions between trapped atoms (or ions) whose separations are relatively well-defined. Simulations analogous to those we have used for random ensembles illustrate the effectiveness of QPM sequences for creating a uniform ensemble of quantum states in DD-coupled atom pairs (e.g. held in optical tweezer arrays) with a narrow spread in $R$ \cite{SM}. 

In the future, QPM may also be employed to suppress microscopic decoherence resulting from time-dependent changes in $V$ for individual elements in an ensemble. These may be caused, for example, by relative thermal motion of atoms or the spatial jitter of a laser beam or trap array. Given the reversal of the phase-advance in successive QPM zones, decoherence caused by temporal variations in $V$ (integrated over a total interaction time $T$) can be minimized through the use of a sufficient number of QPM zones with negligible changes in $V$ during the time $T/N$ spent in each zone \cite{Kutteruf}. 

This work was supported by the NSF.

\bibliography{main}

\begin{thebibliography}{48}
\bibitem [{1()}]{1}%
  \BibitemOpen
  \href@noop {} {}\bibinfo {note} {Population transfer resulting from asymmetric jump sequences in which negligible time is spent at either $\pm$ detuning is indistinguishable from that obtained at constant detuning. The distinctive characteristics of the QPM data are due to the coherent DD-interactions within each zone, not on the transitions between zones.}\BibitemShut {Stop}%
\end{thebibliography}
\end{document}


\title{Supplemental Material}  
\author{C. He and R.R. Jones}     
\affiliation{Department of Physics, University of Virginia, Charlottesville, Virginia 22904-4714} 
\date{\today}
\maketitle

\subsection{Rydberg Atom Excitation, Spatial Distribution, and Detection}
In the experiments, the 300ns, 482nm Rydberg excitation laser pulse has a bandwidth of approximately 3MHz and is produced by chopping a 482nm, $\sim$ 100mW cw laser beam with an acousto-optic modulator. The excitation laser drives atoms from the $^{85}$Rb 5$p_{3/2}$ upper MOT state to the nominal 32$p_{3/2}, |m_j|$=3/2 level in the presence of a static electric field, $F_0\sim $12V/cm. The 5$p \rightarrow$ 32$p_{3/2}, |m_j|$=3/2 optical transition is enabled by weak Stark mixing of the 32$p$ level with neighboring $s$ and $d$ states. The excitation laser is focused to a diameter of ~30$\mu$m through the center of the 0.5-mm diameter MOT with sufficient intensity to saturate the 5$p \rightarrow$ 32$p$ transition at $\simeq$ 50$\%$ excitation probability, creating a cylindrical volume of 32$p_{3/2}, |m_j|=3/2$ atoms. The Rydberg excitation bandwidth is $\sim$ 6MHz, dominated by the natural linewidth of the 5$p_{3/2}$ state. The $^{85}$Rb 32$p$ fine structure splitting is greater than 3 GHz, so there is no laser excitation of 32$p_{1/2}$ atoms \cite{Rb_qd}. Similarly, for $F_0 \sim$ 12 V/cm, the energy splitting between the nominal $32p_{3/2},\abs{m_j}$=1/2 and $32p_{3/2},\abs{m_j}$=3/2 is 150MHz. Thus, there is also negligible excitation of atoms to the $32p_{3/2},\abs{m_j}$=1/2 states.

The atom density in the MOT is calibrated by measuring the fluorescence from the 5$p$ atoms and is controlled by adjusting the background Rb vapor pressure in the MOT vacuum chamber \cite{He}. Prior to the 482nm laser pulse, the atoms in the laser excitation volume are assumed to be randomly distributed with a nearest neighbor distribution function
\begin{equation}
G(R) = 4\pi\rho R^2 e^{-\frac{4}{3}\pi\rho R^3} ~,
\end{equation}
where $R$ is the separation between nearest neighbor atoms and $\rho$ is the atom density \cite{NNeighbor1,NNeighbor2}. This distribution predicts the most probable and average internuclear separations 
\begin{equation}
R_{0}=\sqrt[3]{\frac{1}{2\pi\rho}}
\end{equation}
and 
\begin{equation}
R_{avg}=\sqrt[3]{\frac{3\ln{2}}{4\pi\rho}} \simeq 1.01 R_{0} ~,
\end{equation}
respectively.

Given the relatively low Rydberg principal quantum number, moderate atom densities, and broad Rydberg excitation bandwidths in the experiments, the van der Waals shifts associated with non-resonant Rydberg-Rydberg interactions are negligible for all but the closest atoms in the ensemble. Specifically, the Rydberg blockade radius \cite{blockade} associated with a $\sim$ 6MHz frequency shift is only $\sim$ 2.5$\mu$m. Even at the highest experimental Rydberg density, $\rho$ = 3 $\times$ 10$^9$ atoms/cm$^3$, blockade effects reduce the Rydberg excitation probability in less than 10\% of the atoms in the laser focal volume. Accordingly, the Rydberg atoms can be assumed to be randomly distributed with a saturated Rydberg atom density ($\rho$ = 0.5 - 3.0 $\times$ 10$^9$ atoms/cm$^3$) given by a fixed fraction ($\sim$0.3) of the measured MOT density.

The atoms in the ensemble can be assumed to be frozen throughout the measurements. First, the relative motion of neighboring Rydberg atoms due to DD-forces is negligible over the sub-$\mu$s time scales explored in the experiments. Second, the separation between neighboring Rydberg atom pairs moving with the average relative thermal velocity changes by $<$3$\%$ of the most probable nearest neighbor separation at the highest experimental density. Third, previous measurements have directly shown that for our experimental conditions, the coherence time limits set by relative atom motion are an order of magnitude longer than the maximum times considered in our experiments \cite{Kutteruf}. 

The parallel stainless steel field plates straddling the MOT suppress stray electric fields and reduce inhomogeneities in the applied field to $\simeq$ 0.03\% of the nominal value \cite{He}. Voltage pulses from an arbitrary waveform generator (AWG) applied to the field plates facilitate Stark tuning of the Rydberg atom energies relative to the dipole-dipole (DD) driven 32$p_{3/2},|m_j|$=3/2$\,+\, $32$p_{3/2}, |m_j|$=3/2 $\leftrightarrow$ 32$s\,+\,$33$s$ (i.e. $pp \leftrightarrow ss'$) F{\"o}rster-resonance. At the end of each measurement, a high voltage pulse applied to the field plates ionizes any Rydberg atoms in the interaction region, pushing the resulting ions through a small slot in the field plates toward a microchannel plate (MCP) detector. The time-dependent ion current from the MCP provides a measure of the relative population in different Rydberg states via state-selective field ionization \cite{tfg}. 

\begin{figure}
\centering
\resizebox{80mm}{!}{\includegraphics{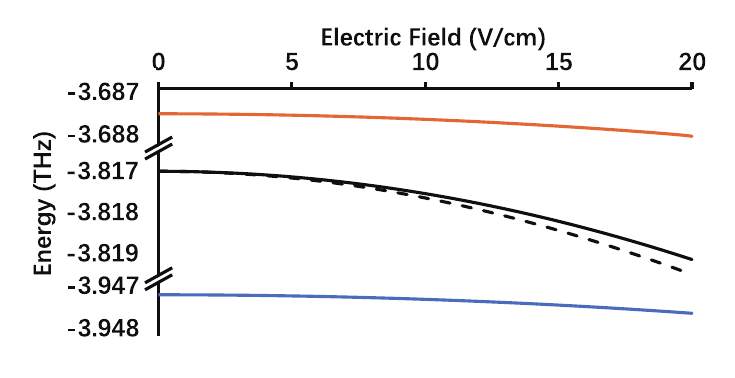}}
\caption{Stark map of $^{85}$Rb $32p_{3/2},\abs{m_j}=3/2$ (black solid), $32p_{3/2},\abs{m_j}=1/2$ (black dash), $32s$ (blue), and $33s$ (orange) single atom binding energies vs electric field.}
\label{fig1}
\end{figure}

\begin{figure}
\centering
\resizebox{80mm}{!}{\includegraphics{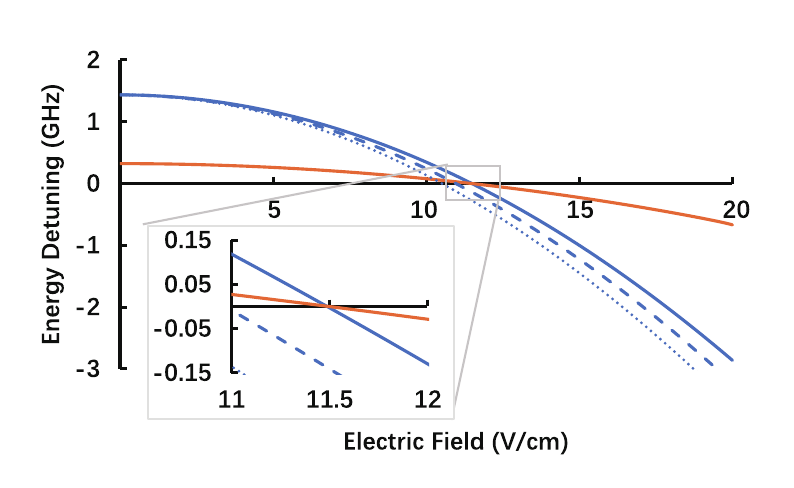}}
\caption{Stark map of the energies of the four degenerate 32$p_{3/2},\abs{m_j}$=3/2 + 32$p_{3/2},\abs{m_j}$=3/2 ($pp$ - blue solid), eight degenerate 32$p_{3/2},\abs{m_j}$=3/2 + 32$p_{3/2},\abs{m_j}$=1/2 (blue dash), four degenerate 32$p_{3/2},\abs{m_j}$=1/2 + 32$p_{3/2},\abs{m_j}$=1/2 (blue dot) and eight degenerate $32s$ + $33s$  ($ss'$ - orange) non-interacting atom pair states vs. electric field. The energies are plotted relative to those of $pp$ and $ss'$ at the $pp \leftrightarrow ss'$ resonance, which occurs in an electric field $F_{0}\simeq$ 11.5 V/cm.}
\label{Stark_pair}
\end{figure}

During a typical measurement of Rydberg population transfer vs. interaction time, 10 independent measurements are averaged for each data point. Five such time scans are averaged in each of the data sets shown in Figs. 2 and 3 of the primary text. The main sources of experimental noise include variations in the number of excited Rydberg atoms (due to statistical fluctuations in the number of atoms within the laser focus and small drifts of the Rydberg excitation laser frequency relative to the 5$p \rightarrow$ 32$p$ transition) as well as small variations ($<$10 mV) in the DC voltage which determines the detuning of atom pairs relative to F{\"o}rster-resonance. 

\subsection{Stark Tuning Near the 32$\mathbf{p_{3/2},|m_j|=}$3/2 + 32$\mathbf{p_{3/2},|m_j|=}$3/2 $\leftrightarrow$ 32$\mathbf{s+}$33$\mathbf{s}$ F{\"o}rster-resonance}

\begin{figure}
\centering
\resizebox{60mm}{!}{\includegraphics{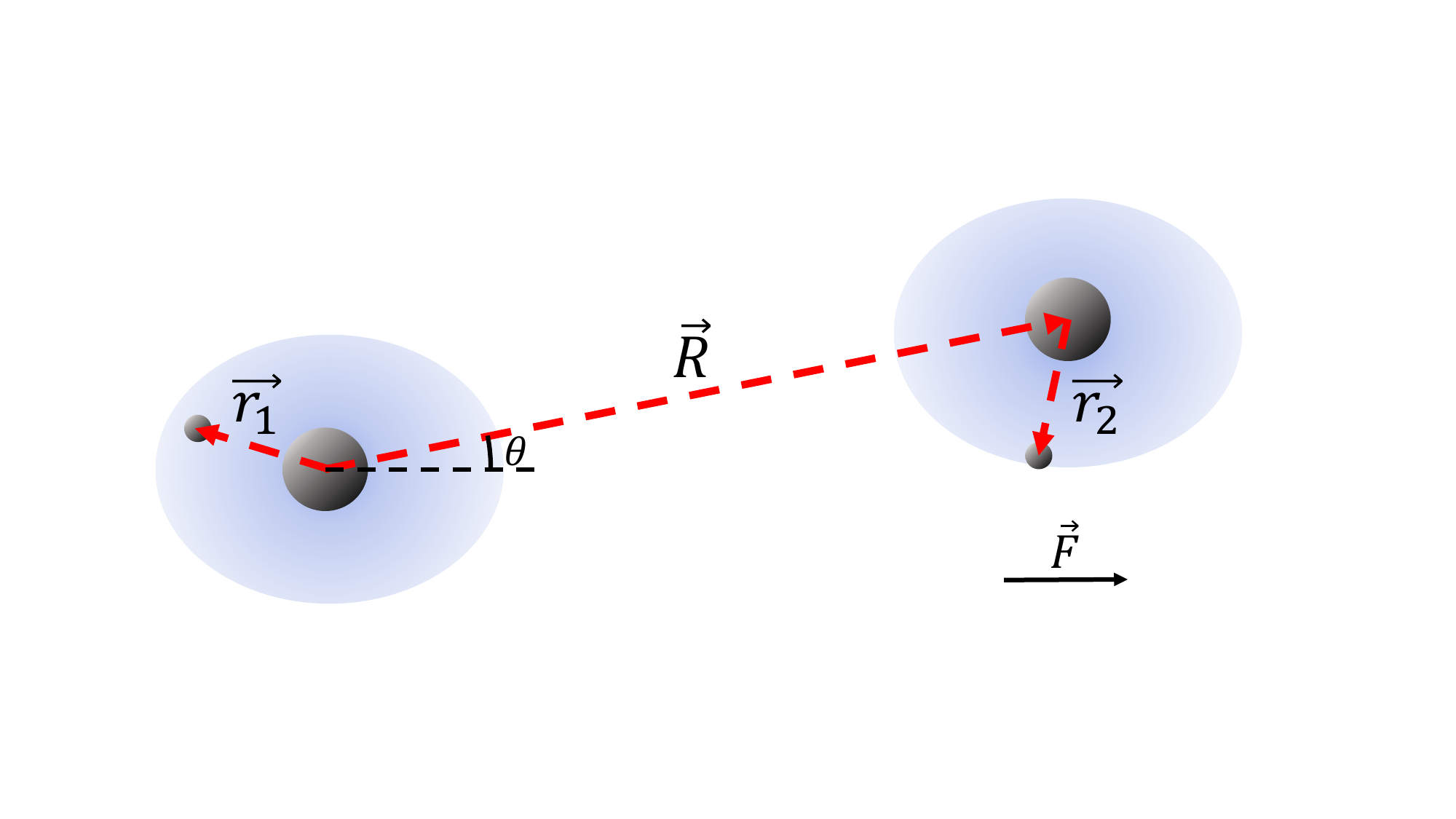}}
\caption{Coordinates relevant to the computation of the DD-interaction between a pair of Rydberg atoms subject to an electric field, $\vec{F} = F\hat{z}$.}
\label{DD_atoms}
\end{figure}

Figure \ref{fig1} shows the calculated binding energies of the nominal  32$p_{3/2}, \abs{m_j}=$1/2, 32$p_{3/2}, \abs{m_j}=$3/2, 32$s$, and 33$s$ Rydberg states in an isolated $^{85}$Rb atom as a function of applied electric field, $F$. In the weak fields of interest in our experiments (11-12V/cm), the eigenstates undergo small quadratic Stark shifts, but principally retain the character of the zero field levels to which they are adiabatically connected. Specifically, the percentages of the zero field levels in these Stark eigenstates are:
32$p_{3/2},\abs{m_j}=1/2$ (98.3$\%$ at 11V/cm and 97.9$\%$ at 12V/cm);
32$p_{3/2},\abs{m_j}=3/2$ (99.0$\%$ at 11V/cm and 98.8$\%$ at 12V/cm);
32$s$ (99.4$\%$ at 11V/cm and 99.3$\%$ at 12V/cm);
33$s$ (99.1$\%$ 11V/cm and 99.0$\%$ 12V/cm). Accordingly, we use zero field state-labels for the Stark eigenstates and the corresponding radial and angular momentum eigenfunctions when calculating matrix elements associated with laser excitation of, and DD-interactions between, the Rydberg atoms. 

The electric field $F_{0}$ at which the $pp \leftrightarrow ss'$ F{\"o}rster-resonance occurs is determined from the energies of ``bare" atom pairs with large internuclear separations $R \rightarrow \infty$, for which atom-atom interactions are negligible. Figure \ref{Stark_pair} shows the relevant $pp$ and $ss'$ pair energies as a function of electric field. The 32$p_{3/2}, \abs{m_j}$=3/2 + 32$p_{3/2}, \abs{m_j}$=3/2 ($pp$) and $32s$+$33s$ ($ss'$) atom pair states are degenerate at $F_{0}$ = 11.49V/cm, defining the resonance field, in good agreement with experiment.   

The energy splitting between the $pp$ and $ss'$ states can be precisely tuned, by varying the external electric field, $F$. Near resonance ($|F-F_{0}| \leq$ 0.5V/cm for our studies), the relative Stark shift of the non-interacting $pp$ and $ss'$ pair states can be approximated by a linear function of $F$. This electric field to resonance ``detuning" conversion function is: $E(F) \simeq$ 170$(F-F_{0}) \frac{\text{MHz}}{(\text{V/cm})}$. 

\subsection{Dipole-Dipole Interaction Between Two Well-Separated Rydberg Atoms}
Consider a pair of Rydberg atoms, each containing a positively charged ion core and a highly excited electron. For ion separations $R$ much greater than the spatial extent of the relevant Rydberg wavefunctions on the two atoms, the lowest order term in the neutral atom-atom multipole coupling is dipole-dipole (DD). The DD interaction between two atoms is \cite{tfg,Robicheaux_DD}

\begin{equation}
    V_{DD}=\frac{\vec{r_1}\cdot\vec{r_2}}{R^3}-\frac{3(\vec{r_1}\cdot\vec{R})(\vec{r_2}\cdot\vec{R})}{R^5},
\label{DD_full}
\end{equation}
where $\vec{r_1},\vec{r_2}$ are the positions of electrons 1 and 2 relative to their respective ions, and $\vec{R}$ is the position of ion 2 relative to 1 (Fig. \ref{DD_atoms}). 

We define $\hat{z}$ as the direction of an external electric field $\vec{F}$ in the lab frame, and express Eq. \ref{DD_full} using lab frame coordinates with $\vec{r_1}=(x_1,y_1,z_1), \vec{r_2}=(x_2,y_2,z_2) $ and $\vec{R}=(X,Y,Z)$. The internuclear distance is $R=\sqrt{X^2+Y^2+Z^2}$, and the direction of $\vec{R}$ can be defined using polar and azimuthal angles $\Theta=\text{arctan}\frac{\sqrt{X^2+Y^2}}{Z}$ and $\Phi=\text{arctan}\frac{Y}{X}$, respectively. This gives
\begin{equation}
\begin{split}
    V_{DD}=\frac{1}{R^3}\big[&(x_1x_2+y_1y_2+z_1z_2)\\
    &-3(x_1\text{sin}\Theta\text{cos}\Phi+y_1\text{sin}\Theta\text{sin}\Phi+z_1\text{cos}\Theta)\\
    &\cdot(x_2\text{sin}\Theta\text{cos}\Phi+y_2\text{sin}\Theta\text{sin}\Phi+z_2\text{cos}\Theta)\big],
\end{split}
\label{DD_lab}
\end{equation}
in the lab frame. Although it is not immediately obvious from Eq. \ref{DD_lab}, for a single pair of atoms, $V_{DD}$ is independent of the azimuthal angle $\Phi$ (describing rotations of $\vec{R}$ about the electric field axis, $\hat{z}$). However, the independence no longer holds for ensembles with more than two atoms where beyond nearest neighbor interactions play a role.

Due to the substantial mass of the ions, and their relatively large separation, $\vec{R}$ can be treated as a classical parameter. However, the positions of the Rydberg electrons, $x_1,y_1,z_1,x_2,y_2,z_2$ must be treated using quantum mechanical operators. We exploit the symmetry of the electronic angular momentum eigenstates to reduce the complexity of the computation of $V_{DD}$, writing the Cartesian position operators for the two Rydberg electrons in terms of spherical tensor operators \cite{Racah,Edmonds}, $x_{1,2}=\frac{1}{\sqrt{2}}r_{1,2}\left(C^{(1)}_{-1}-C^{(1)}_{1}\right),y_{1,2}=\frac{i}{\sqrt{2}}r_{1,2}\left(C^{(1)}_{-1}+C^{(1)}_{1}\right),z_{1,2}=r_{1,2} C^{(1)}_{0}$, where $r_{1,2}$ are the radial coordinate operators for the two electrons:

\begin{equation}
\begin{split}
    V_{DD}=\frac{r_{1} r_{2}}{R^3}\bigg[&-\frac{3}{2}C^{(1)}_{1}C^{(1)}_{1}\text{sin}^2\Theta(\text{cos}2\Phi-i\text{sin}2\Phi) \\
    &+\frac{3}{2\sqrt{2}}\left(C^{(1)}_{1}C^{(1)}_{0}+C^{(1)}_{0}C^{(1)}_{1}\right)\\
    &\cdot\text{sin}2\Theta(\text{cos}\Phi-i\text{sin}\Phi) \\
    &+C^{(1)}_{0}C^{(1)}_{0}(1-3\text{cos}^2\Theta) \\
    &+\left(C^{(1)}_{1}C^{(1)}_{-1}+C^{(1)}_{-1}C^{(1)}_{1}\right)\left(\frac{3}{2}\text{sin}^2\Theta-1\right) \\
    &-\frac{3}{2\sqrt{2}}\left(C^{(1)}_{-1}C^{(1)}_{0}+C^{(1)}_{0}C^{(1)}_{-1}\right)\\
    &\cdot\text{sin}2\Theta(\text{cos}\Phi+i\text{sin}\Phi) \\
    &-\frac{3}{2}C^{(1)}_{-1}C^{(1)}_{-1}\text{sin}^2\Theta(\text{cos}2\Phi+i\text{sin}2\Phi)\bigg],
\end{split}
\label{DD_full_C}
\end{equation}
where, in each product of two spherical tensor operators, the operator on the left acts on the Rydberg electron in atom 1 and the operator on the right acts on the Rydberg electron in atom 2. Notably, the projection of total electronic angular momentum along the electric field axis, $M_J$, is not conserved in the presence of both the static field and $V_{DD}$. $V_{DD}$ can couple states for which: i) $M_J$ changes by 0, $\pm1$, or $\pm2$; and ii) the angular momentum quantum numbers $\ell_1$ and $\ell_2$ for the individual electrons each change by $\pm1$.

\subsection{Pair Eigenstates and Energies Near the 32$\mathbf{p_{3/2},|m_j|=}$3/2 + 32$\mathbf{p_{3/2},|m_j|=}$3/2 $\leftrightarrow$ 32$\mathbf{s+}$33$\mathbf{s}$ F{\"o}rster-resonance}

The DD-coupling between atoms alters the bare Rydberg pair states and the energies that are shown in Figs. \ref{fig1} and \ref{Stark_pair}. The effects are small when the magnitude of the energy separation between the interacting states $|E|$ is much greater than the magnitude of the relevant matrix element of $V_{DD}$ between them. The prefactor, $\frac{r_1 r_2}{R^3}$, in $V_{DD}$ sets the upper scale for those matrix elements, since the angular terms are of order unity.  In general, for a given Rydberg state with principal quantum number $n$, the radial matrix elements connecting it to other levels are largest when both states have very nearly the same effective principal quantum number $\nu = n-\delta$, where $\delta$ is the quantum defect. In this case, the matrix elements scale as $\nu^2$. Their magnitude decreases substantially for differences in $\nu$ greater than 1 \cite{tfg}.

Our experiments focus on the effect of DD-interactions on initially excited $^{85}$Rb 32$p$ ($\nu \simeq $29.4) atoms. By far the largest relevant matrix elements of $V_{DD}$ involve the 32$s$ or 33$s$ states ($\nu \simeq$ 28.9 or 29.9, respectively) or the 30$d$ or 31$d$ levels ($\nu \simeq$ 28.7 or 29.7, respectively) \cite{Rb_qd}. For two Rb atoms separated by the most probable nearest neighbor distance $R_0$ in a random ensemble at density $\rho$, those matrix elements have maximum magnitudes on the order of $2 \pi \rho \nu^4$, roughly 5MHz at a typical experimental Rydberg density, $\rho \sim 10^9$ atoms/cm$^3$. However, all Rydberg atom pair combinations, except 32$s$33$s$, are separated in energy from 32$p$32$p$ by several cm$^{-1}$ or more over the range of applied fields we consider, greater than the largest DD matrix elements by at least five orders of magnitude. Therefore, we can safely neglect all pair combinations other than 32$p$32$p$ and 32$s$33$s$ when considering the DD-driven dynamics of an ensemble of initially excited 32$p$ atoms. Similarly, given that both the maximum DD matrix elements and the Rydberg excitation bandwidth have magnitudes of several MHz, the 3.4GHz 32$p$ fine-structure splitting ensures that 32$p_{1/2}$ states play no role in the dynamics of isolated atom pairs \cite{Rb_qd}. Thus, the bare pair basis states that are potentially relevant to the experiments are the 24 (sixteen 32$p_{3/2}$ + 32$p_{3/2}$ and eight $ss'$ levels) whose field-dependent energies are shown in Fig. \ref{Stark_pair}.

\begin{figure}
\centering
\resizebox{80mm}{!}{\includegraphics{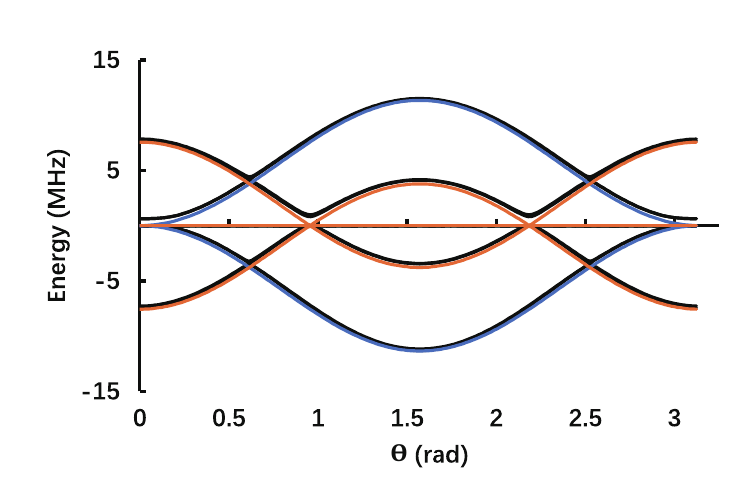}}
\caption{On-resonance pair eigenenergies vs alignment angle $\Theta$ in a static field $F = F_0$, for an atom separation equal to the average nearest neighbor separation for a random ensemble with a Rydberg density of $3\times10^9$/cm$^3$. The energies are plotted relative to the bare 32$p_{3/2},|m_j|$=3/2+32$p_{3/2},|m_j|$=3/2 level. The solid black curves show the results from the full 24 state basis (i.e. including 32$p_{3/2},|m_j|$=1/2 states). The blue and orange curves are from the reduced basis calculations (excluding 32$p_{3/2},|m_j|$=1/2 states). The blue lines show eigenstates that are linear combinations of bare angular momentum states with $\abs{M_J}$=1 (for $ss'$) and $\abs{M_J}$=3 (for $pp$). The orange lines show eigenstates that are linear combinations of $pp$ and $ss'$ with $\abs{M_J}$=0. The lines at zero energy (only orange is visible) are states that are antisymmetric linear combinations, $\ket{ss'}_A= (\ket{ss'}-\ket{s's})/\sqrt{2}$, which do not couple to $pp$.
}
\label{3.a}
\end{figure}

Due to the differential Stark shift, in electric fields near the 32$p_{3/2},|m_j|$=3/2+32$p_{3/2},|m_j|$=3/2 $\leftrightarrow$ 32$s+$33$s$ (i.e. $pp \leftrightarrow ss'$) F{\"o}rster-resonance, the pair states involving 32$p_{3/2},|m_j|$=3/2+32$p_{3/2},|m_j|$=1/2 and 32$p_{3/2},|m_j|$=1/2+32$p_{3/2},|m_j|$=1/2 are split from the 32$p_{3/2},|m_j|$=3/2+32$p_{3/2},|m_j|$=3/2 levels by $\sim$140MHz and $\sim$280MHz, respectively (see Fig. \ref{Stark_pair}). Accordingly, there is negligible direct laser excitation of atom pairs involving 32$p_{3/2},|m_j|$=1/2, and the DD couplings between the initially populated 32$p_{3/2},|m_j|$=3/2+32$p_{3/2},|m_j|$=3/2 pairs and the 32$p_{3/2},|m_j|$=3/2+32$p_{3/2},|m_j|$=1/2 or 32$p_{3/2},|m_j|$=1/2+32$p_{3/2},|m_j|$=1/2 levels are expected to be weak. 

Fig. \ref{3.a} confirms that expectation, comparing two calculations of the pair eigenenergies in the presence of the DD interaction at resonance ($F = F_0$), as a function of the orientation angle $\Theta$. The solid black lines show the energies computed using the 24 state basis (i.e. including 32$p_{3/2},|m_j|=$1/2 states). The orange and blue curves are the energies computed using a reduced basis with only 32$p_{3/2},|m_j|=$3/2, 32$s$ and 33$s$ states. The energies are plotted relative to the bare 32$p_{3/2},|m_j|=$3/2 + 32$p_{3/2},|m_j|=$3/2 level. As expected, there are only minor differences in the results of the two calculations. The eigenenergies agree to better than 1MHz, with the largest differences at angles where the DD-coupling is zero in the reduced basis. The 32$p_{3/2},|m_j|$=1/2 character in the eigenstates is also quite small. Accordingly, starting from a pure 32$p_{3/2},|m_j|=$3/2 ensemble, for short interaction times ($<$ 1$\mu$s) the DD-driven dynamics of atom pairs are accurately described using the reduced basis. Over these brief timescales, spontaneous emission, transitions between Rydberg levels due to black-body radiation, and thermal and/or DD-interaction-induced atom motion can all be reasonably neglected. 

\subsection{Approximate 2-Level DD-Coupled Systems}
A distinct advantage of the reduced pair-state basis is that the level structure and dynamics collapse to that of four independent, coupled two-level systems, distinguished by the signs of the projections of total electronic angular momentum for the individual 32$p_{3/2},|m_j|=$3/2 electrons in the initially excited $pp$ state. Each of those distinct $pp$ basis states is coupled to a single symmetric linear combination of $ss'$ and $s's$ states, $\ket{ss'}_S= (\ket{ss'}+\ket{s's})/\sqrt{2}$, which are also distinguished by the $m_j$ values of the two electrons. Four additional basis states that are antisymmetric linear combinations $\ket{ss'}_A= (\ket{ss'}-\ket{s's})/\sqrt{2}$ with $(m_{j},m_{j'})=$(1/2,1/2),(-1/2,-1/2),(1/2,-1/2) and (-1/2,1/2), do not couple to $pp$ and, therefore, remain unpopulated and not relevant to the dynamics. The basis vectors and corresponding DD matrix elements for the four independent two-level systems are detailed below.

\begin{equation*}
\begin{pmatrix} \ket{pp} \\[2pt] \ket{ss'}_S \end{pmatrix} _{++} \; =
\begin{pmatrix} pp(3/2,3/2) \\[2pt] \frac{1}{\sqrt{2}}[ss'(1/2,1/2)+s's(1/2,1/2)] \end{pmatrix}\;
\end{equation*}
\begin{equation}
V_{++} = -\frac{r_{ps}r_{ps'}}{\sqrt{2} R^3} \text{sin}^2\Theta(\text{cos}2\Phi-i\text{sin}2\Phi)
\label{V++}
\end{equation}

\begin{equation*}
\begin{pmatrix} \ket{pp} \\[2pt] \ket{ss'}_S \end{pmatrix} _{--} \; =
\begin{pmatrix} pp(-3/2,-3/2) \\[2pt] \frac{1}{\sqrt{2}}[ss'(-1/2,-1/2)+s's(-1/2,-1/2)] \end{pmatrix}\;
\end{equation*}
\begin{equation}
V_{--} = V_{++}^*
\end{equation}

\begin{equation*}
\begin{pmatrix} \ket{pp} \\[2pt] \ket{ss'}_S \end{pmatrix} _{+-} \; =
\begin{pmatrix} pp(3/2,-3/2) \\[2pt] \frac{1}{\sqrt{2}}[ss'(1/2,-1/2)+s's(1/2,-1/2)]
\end{pmatrix}\;
\end{equation*}
\begin{equation}
V_{+-} = \frac{r_{ps}r_{ps'}}{\sqrt{2} R^3} \left(\text{sin}^2\Theta - \frac{2}{3} \right)
\end{equation}

\begin{equation*}
\begin{pmatrix} \ket{pp} \\[2pt] \ket{ss'}_S \end{pmatrix} _{-+} \; =
\begin{pmatrix} pp(-3/2,3/2) \\[2pt] \frac{1}{\sqrt{2}} [ss'(-1/2,1/2)+s's(-1/2,1/2)]\end{pmatrix}\;
\end{equation*}
\begin{equation}
V_{-+} = V_{+-}
\label{V-+}
\end{equation}
The radial matrix elements $r_{ps}$ =964 and $r_{ps'}$=941 are computed numerically. 

For each 2-level system, with a detuning $E$ between the $pp$ and $ss'$ basis states, the Hamiltonian takes the simple form:
\begin{equation}
    H=
\begin{pmatrix}
0 & V \\
V^* & E \\
\end{pmatrix},
\end{equation}
with eigenvalues
\begin{equation}
E_{\pm} = \frac{1}{2}\Big(E\pm\sqrt{E^2+4\abs{V}^2}\Big) 
\end{equation}
for $V$ = $V_{++}$, $V_{--}$, $V_{+-}$, or $V_{-+}$ depending on the initial $pp$ state. By inspection, and as expected from the symmetry of the two-atom geometry, the eigenvalues are independent of $\Phi$ for any of the four possible values of $V$. Thus without loss of generality, we can set $\Phi$ = 0, ensuring that $V$ is real. The relevant eigenstates corresponding to the respective eigenvalues $E_+$ and $E_-$ can then be written
\begin{equation*}
\ket{+}=\cos{\frac{\alpha}{2}}\ket{pp}+\sin{\frac{\alpha}{2}}\ket{ss'}_S
\end{equation*}
\begin{equation}
\ket{-}=\cos{\frac{\alpha}{2}}\ket{ss'}_S-\sin{\frac{\alpha}{2}}\ket{pp}
\end{equation}
where $\tan{\alpha}=\frac{2V}{E}$. The eigenvalues and eigenstates for $V_{++}$ and $V_{--}$ ($pp, ~\abs{M_J} = 3$ and $ss', ~\abs{M_J} = 1$) are identical, as they are for $V_{+-}$ and $V_{-+}$ ($pp, ~\abs{M_J} = 0$ and $ss', ~\abs{M_J} = 0$). For a random ensemble with a density $\rho$ = 10$^{9}$ cm$^{-3}$, the angle averaged DD-matrix elements for an atom pair separated by the most probable nearest neighbor distance $R_0$ are: $\langle V_{++,--} \rangle \simeq 2.0$MHz and $\langle V_{+-,-+} \rangle \simeq 0.66$MHz. 

\begin{figure}
\centering
\resizebox{75mm}{!}{\includegraphics{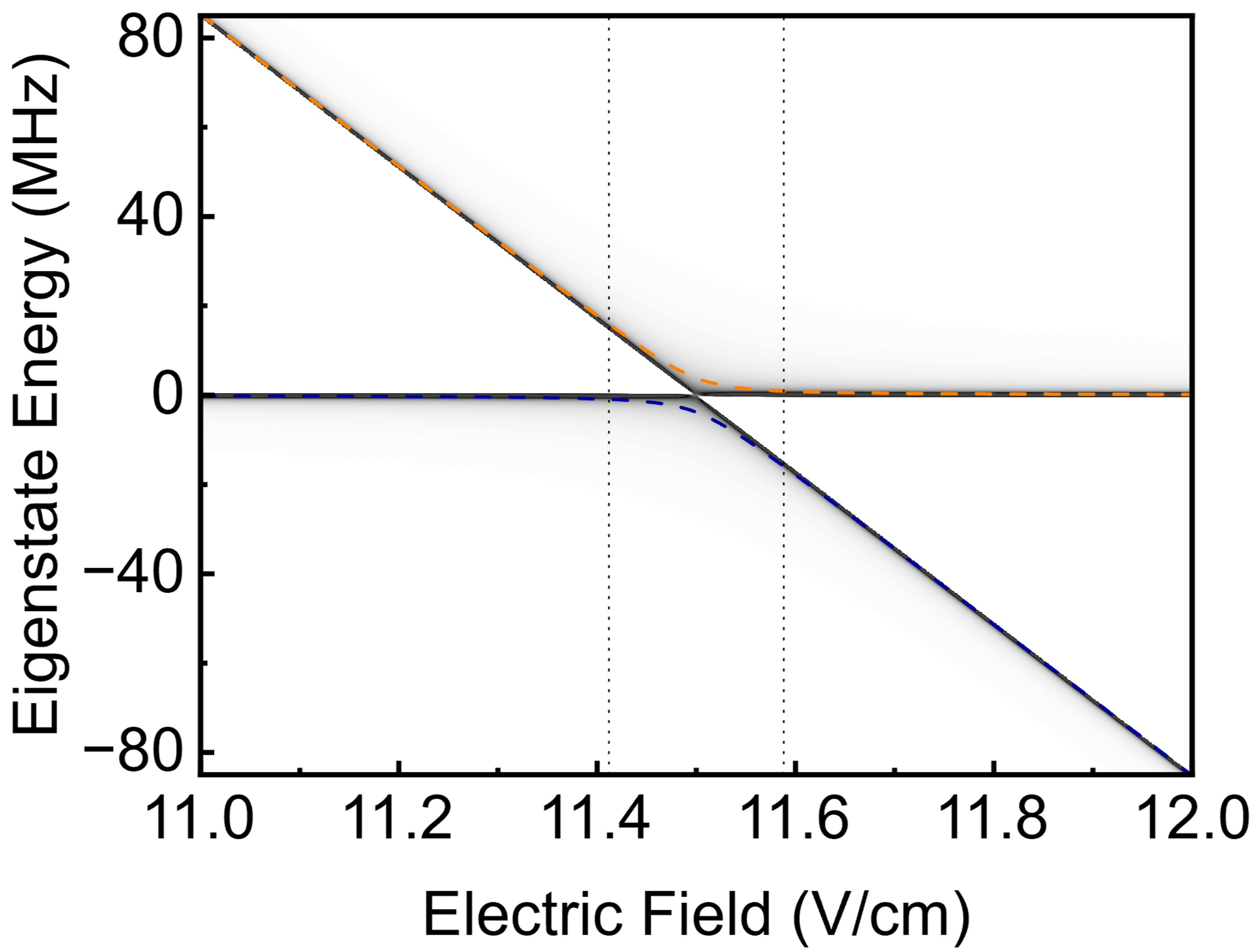}}
\caption{Distribution of calculated energies for a random ensemble of DD-coupled $pp$ and $ss'$ atom pairs vs electric field. The distribution of atom separations corresponds to that for nearest neighbor distances at a density $\rho =$ 2 $\times$ 10$^{9}$ cm$^{-3}$. The $pp$ atom pairs are assumed to have equal probabilities for total electronic angular momentum projections, $\abs{M}=3,0$. The gray scale reflects the number of energy curves passing through each point. The dashed orange and blue lines show the eigenenergies for a pair of atoms with the average $V$ for the random ensemble. The vertical dotted lines show typical fields used for the quasi-phase-matching sequences described in the main text. }
\label{LZ}
\end{figure}

For a given pair of atoms with separation $R$ and orientation angle $\Theta$, the eigenstates form an avoided level crossing as a function of the detuning $E$ (or the applied field $F$) with a minimum curve separation of 2$V(R,\Theta)$ at $E$=0 (i.e. at $F=F_0\simeq$ 11.5 V/cm). Fig. \ref{LZ} shows the distribution of calculated avoided level crossings for a random ensemble at a density $\rho =$ 2 $\times$ 10$^{9}$ cm$^{-3}$. The gray scale reflects the number of energy curves passing through each point. The dashed orange and blue lines show the energy levels for an atom pair with the ensemble averaged value of $V$.  

\subsection{Dynamics of the 2-Level DD-Coupled Systems}
\begin{figure}
\centering
\resizebox{85mm}{!}{\includegraphics{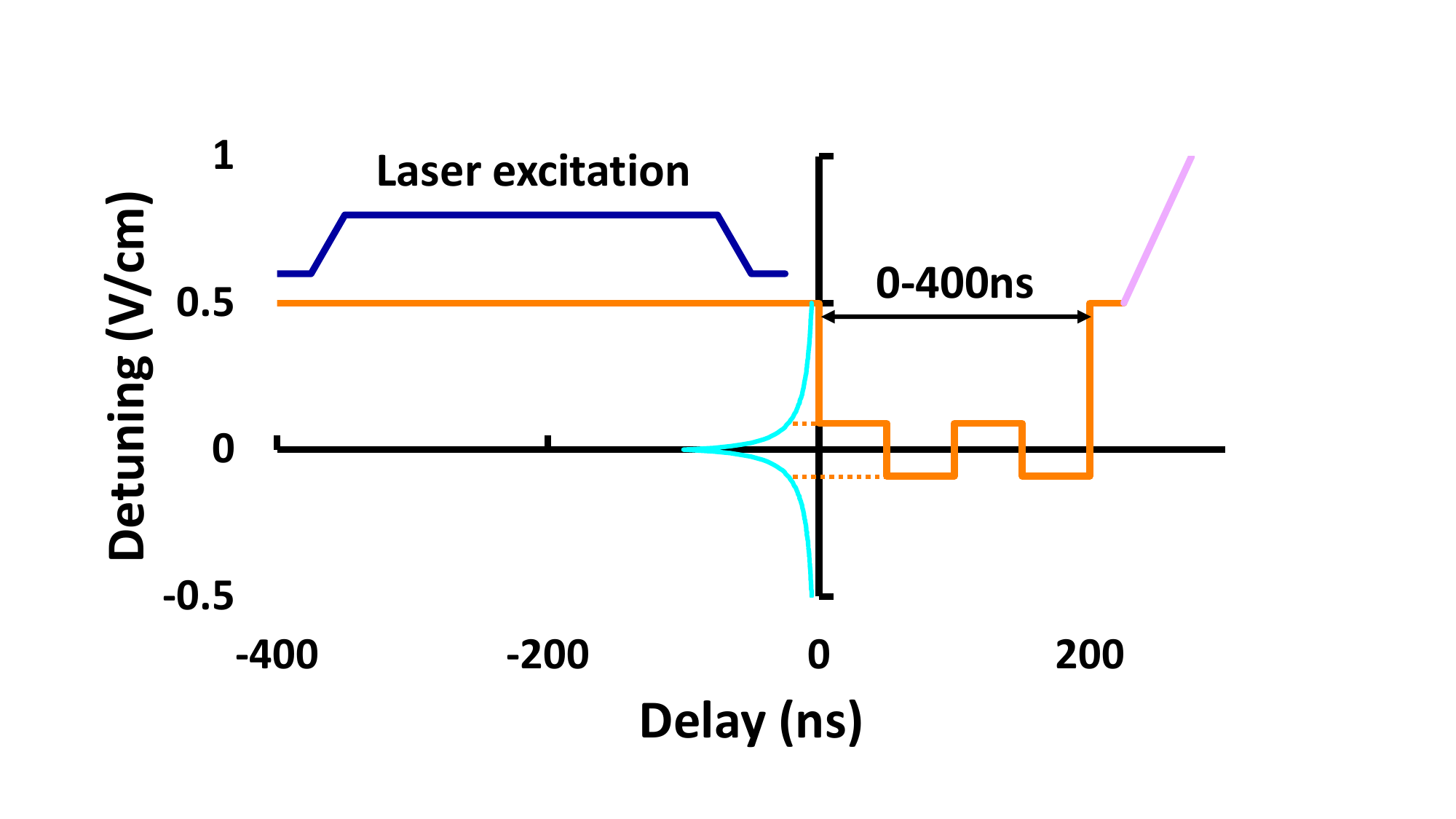}}
\caption{Schematic of the experimental timing including laser excitation of the 32$p$ Rydberg atoms (dark blue line), control of the DD-interaction through tuning-field steps (orange line) relative to the resonance lineshape (cyan line), and field ionization detection (magenta line).}
\label{timing}
\end{figure}

In the experiments, all of the Rydberg atoms in the ensemble are initially excited to 32$p_{3/2},|m_j|=$3/2 states in an electric field, $F$=12 V/cm, such that they are well-detuned from resonance (see Fig. \ref{LZ}). As shown in Fig. \ref{timing}, at $t=0$ the atoms are tuned on, or near, resonance using a voltage step or a sequence of voltage steps with fast ($\simeq$ 2ns) rise (or fall) times, allowing $pp$ atoms to interact via the DD-interaction. Notably, the population transfer resulting from asymmetric jump sequences in which negligible time is spent at either positive or negative detuning is indistinguishable from that obtained at constant detuning. So, the distinctive characteristics of the QPM data are due to the coherent DD-interactions within each zone, not on the transitions between zones \cite{nonzerojump}. After a total interaction time $T$ the atoms are decoupled by tuning them very far from resonance and then field ionized to measure the population transfer probability from $pp$ to $ss'$. Within the nearest neighbor approximation, the dynamics of each atom pair can be described using an analytic expression. 

If the detuning $E$ is constant throughout the interaction time, each initial pair $\Psi(0)= \ket{pp}$ evolves to
\begin{equation}
    \Psi(T) = C_{pp}(T) \ket{pp} + C_{ss'}(T) \ket{ss'}_S
\end{equation}
where the amplitudes $C_{pp}(T)$ and $C_{ss'}(T)$ are given by the standard transformation \cite{Meystre}:

\begin{equation}
\begin{split}
\begin{pmatrix} C_{pp}(T) \\[4pt] C_{ss'}(T) \end{pmatrix} \; = 
U(E,T) \begin{pmatrix} C_{pp}(0) \\[4pt] C_{ss'}(0) \end{pmatrix}
\end{split}
\end{equation}
with $C_{pp}(0)$= 1, $C_{ss'}(0)$=0 and 
\begin{equation}
    U(E,T) = \begin{pmatrix} \cos{\frac{\phi}{2}}-i \frac{E}{\Gamma} \sin{\frac{\phi}{2}} & i \frac{2 V}{\Gamma} \sin{\frac{\phi}{2}} \\[7pt]
    i \frac{2 V}{\Gamma} \sin{\frac{\phi}{2}} & \cos{\frac{\phi}{2}}+i \frac{E}{\Gamma} \sin{\frac{\phi}{2}} \end{pmatrix}
\end{equation}
where $\phi = \Gamma T$ is the Rabi phase and $\Gamma=\sqrt{E^2 + 4V^2}$ is the generalized Rabi frequency.

Accordingly, the population transfer probability to $ss'$ is
\begin{equation}
    \abs{C_{ss'}(T)}^2 = \frac{4V^2}{\Gamma^2} \sin^2{\frac{\phi}{2}} ~,
    \label{Rabi}
\end{equation}
exhibiting Rabi oscillations as a function of $T$. The maximum population transfer probability has a Lorentzian lineshape as a function of detuning, with a full-width at half-maximum (FWHM) $\Delta E = 4V$ (see Fig. 1 in the main text).

\subsection{Dynamics During QPM Sequences}
To implement the time-domain quasi-phase matching (QPM) sequences described in the main text, the sign of the detuning is rapidly reversed at the start of $N$ consecutive time zones, with the atoms spending a time $T/N$ at a constant detuning $\pm E$ in each zone. In this case, the system evolution is described by a product of constant detuning transformations
\begin{equation}
\begin{split}
\begin{pmatrix} C_{pp}(T) \\[4pt] C_{ss'}(T) \end{pmatrix} \; = 
U_{QPM,N} \begin{pmatrix} C_{pp}(0) \\[4pt] C_{ss'}(0) \end{pmatrix}
\end{split}
\end{equation}
with
\begin{equation}
    U_{QPM,N} = [U(-E,T/N) \cdot U(E,T/N)]^{N/2}
\end{equation}
\noindent
for an even number of zones. Explicitly, for $N$=2 zones, $U_{QPM,2} = [U(-E,T/2) \cdot U(E,T/2)]=$ \newline

\noindent
\begin{small}
$\begin{pmatrix} \cos^2{\frac{\phi}{4}}+ \frac{E^2-4V^2}{\Gamma^2} \sin^2{\frac{\phi}{4}} &  -\frac{4EV}{\Gamma^2} \sin^2{\frac{\phi}{4}}+i \frac{4V}{\Gamma} \sin{\frac{\phi}{4}}\cos{\frac{\phi}{4}}\\[7pt]
  \frac{4EV}{\Gamma^2} \sin^2{\frac{\phi}{4}}+i \frac{4V}{\Gamma} \sin{\frac{\phi}{4}}\cos{\frac{\phi}{4}} & \cos^2{\frac{\phi}{4}}+ \frac{E^2-4V^2}{\Gamma^2} \sin^2{\frac{\phi}{4}}
\end{pmatrix}$
\end{small}
\bigskip

For large detunings $E \gg V$, we have $\frac{E}{\Gamma} \simeq 1$ and $\frac{2V}{\Gamma} \ll 1$. In this case, it is straightforward to show that the $N$-zone QPM transfer matrix can be approximated as
\begin{equation}
U_{QPM,N} \approx
 \begin{pmatrix} 1 & i \frac{2V}{\Gamma_N}\sin{\frac{\phi_N}{2}}e^{i \frac{\phi_N}{2}} \\[4pt]  i \frac{2V}{\Gamma_N}\sin{\frac{\phi_N}{2}} e^{-i \frac{\phi_N}{2}} & 1 \end{pmatrix}
\end{equation}
for $\Gamma_N = \Gamma / N$ and $\phi_N = \Gamma_N T = \phi/N$, provided $2^{N/2} \left( \frac{2V}{\Gamma_N}\right)^2 \sin^2{\frac{\phi_N}{2}} \ll 1$ so that the net population transfer to $ss'$ 
\begin{equation}
    \abs{C_{ss'}(T)}^2 = \frac{4V^2}{\Gamma_N^2} \sin^2{\frac{\phi_N}{2}} ~,
    \label{Rabi_N}
\end{equation}
is small.

This expression is identical to that derived for constant detuning (Eq. \ref{Rabi}), except that the generalized Rabi frequency $\Gamma$ has been replaced by $\Gamma_N$. In terms of population transfer, the net result of the QPM sequence is an increase in the period of the off-resonant Rabi oscillations by a factor of $N$ and an increase in the amplitude of those oscillations by a factor of $N^2$. As discussed in the main text, the enhancement of the maximum population transfer afforded by QPM enables the creation of more uniform quantum states throughout an ensemble through the use of far off-resonant interactions for which the variations in the generalized Rabi frequency are substantially reduced. Notably, as seen in the experiments, for large population transfers where the condition $2^{N/2} \left( \frac{2V}{\Gamma_N}\right)^2 \sin^2{\frac{\phi_N}{2}} \ll 1$ no longer holds, depletion of the $pp$ amplitude in each atom pair limits the probability amplitude that is available for transfer during some number of zones. Accordingly, the QPM enhancement factor is less than $N^2$ and the effective Rabi oscillations are no longer strictly sinusoidal. 

\begin{figure*}
\centering
\resizebox{0.7\textwidth}{!}{\includegraphics{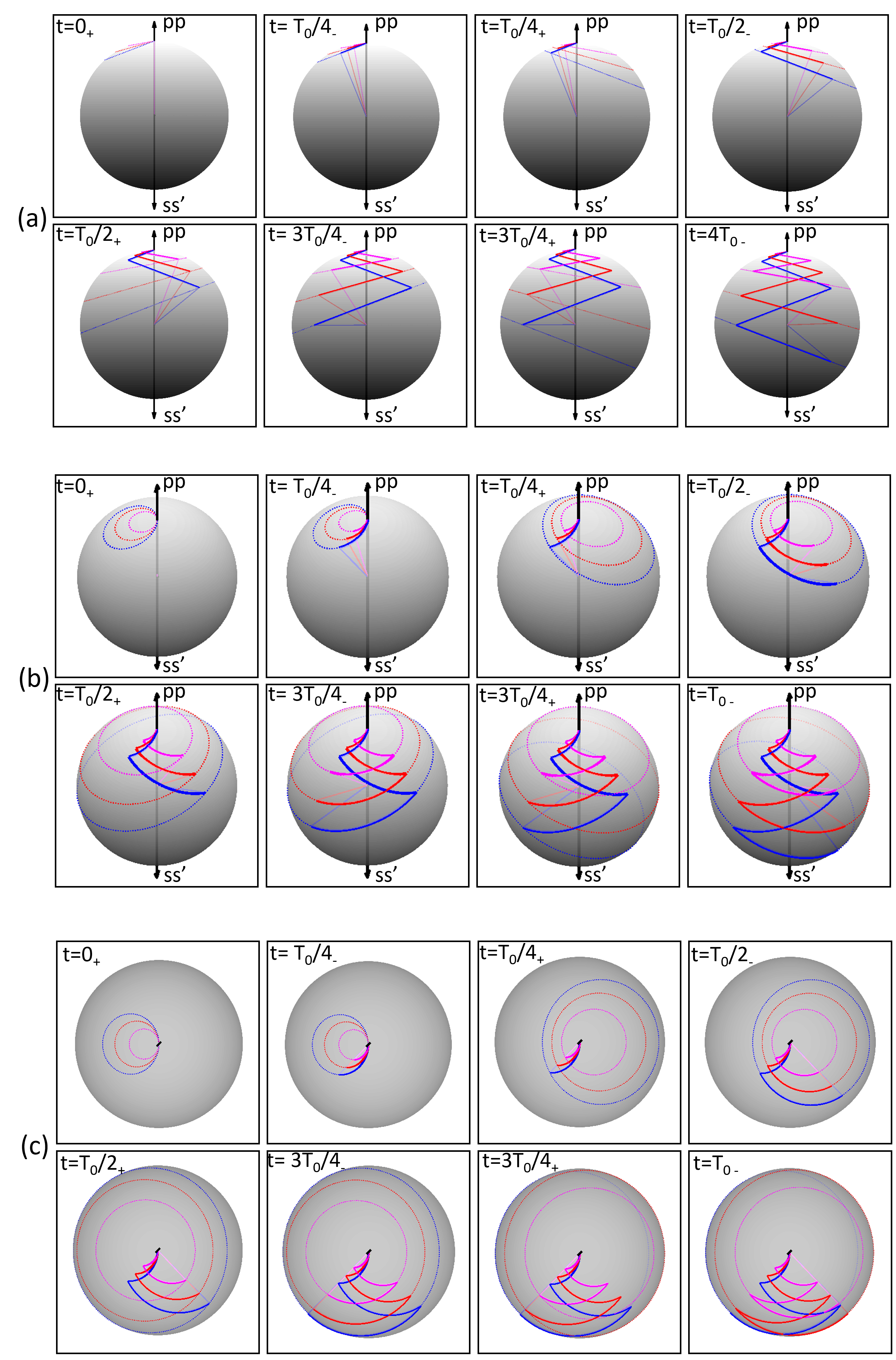}}
\caption{Bloch sphere representation of the evolution of the state vectors for 3 atom pairs subject to different DD-interaction strengths ($V$ = 0.09, 0.14 and 0.19 for magenta, red and blue, respectively) at the same detuning from resonance, $E$ = 0.99, for a $N$ = 4 time-zone QPM sequence. Groups (a), (b), and (c) show identical evolution, but from different viewing angles. The time during the evolution is noted in the upper left corner of each of the eight panes within the three groups. The subscripts on the time labels indicate that the snapshot corresponds to conditions just before ($-$) or just after ($+$) the detuning reversal at the interface between two time zones. The time spent at constant detuning in each zone is approximately one-quarter of a Rabi cycle, $T_0 /4 = \pi/2$. The thin dotted lines that form circles on the surface of the sphere show the projected paths of the state vectors at the current detuning. Thick solid lines show the paths actually traversed on the Bloch sphere up to that point in time. Thin solid radial lines show the locations of the Bloch vectors at that instant. Note that due to the reversal of the direction of the Bloch vector rotation, the small differences in the azimuthal phase angles due to the different coupling strengths do not grow in time, as their magnitudes at the end of each time zone are limited to those at the end of the first zone.    
}
\label{QPM1}
\end{figure*}

\begin{figure*}
\centering
\resizebox{0.82\textwidth}{!}{\includegraphics{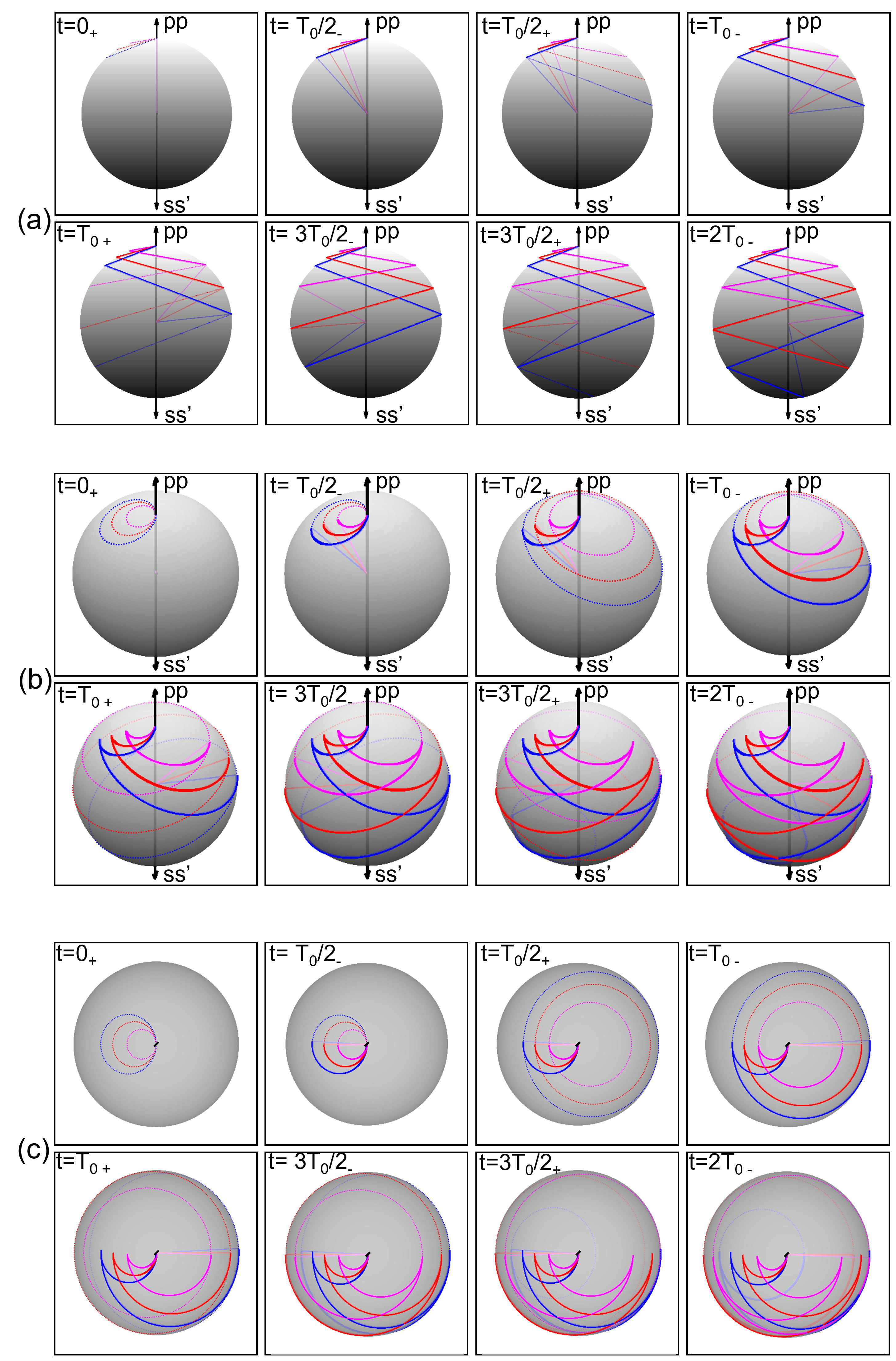}}
\caption{Bloch sphere representation of the evolution of the state vectors for 3 atom pairs subject to a 4-zone QPM sequence, analogous to Fig. \ref{QPM1} except that the time spent in each zone is approximately one-half of a Rabi cycle, $T_0 /2 = \pi$.
}
\label{QPM2}
\end{figure*}

\begin{figure*}
\centering
\resizebox{0.82\textwidth}{!}{\includegraphics{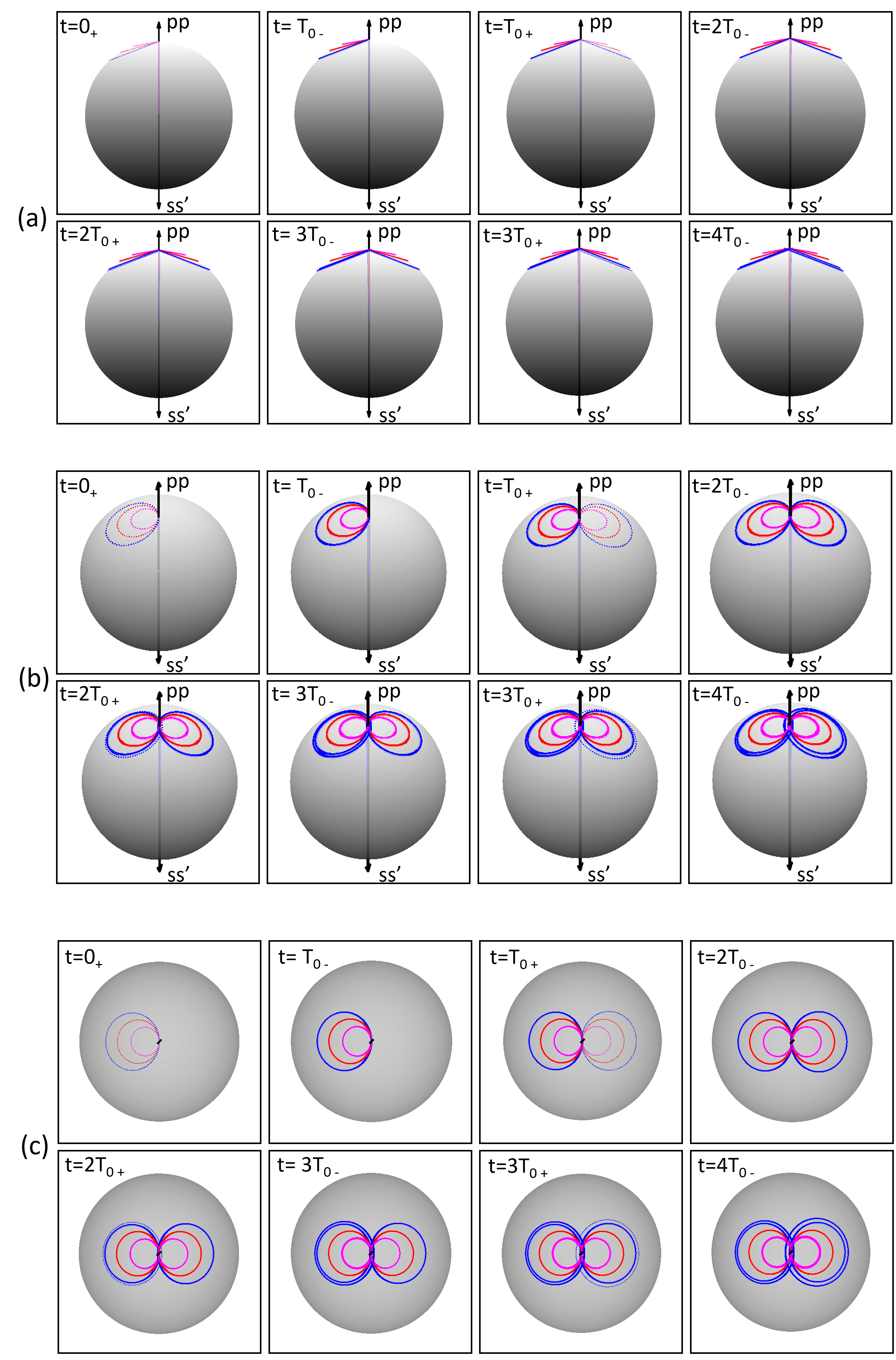}}
\caption{Bloch sphere representation of the evolution of the state vectors for 3 atom pairs subject to a 4-zone QPM sequence, analogous to Fig. \ref{QPM1} except that the time spent in each zone is approximately one full Rabi cycle, $T_0 = 2 \pi$.
}
\label{QPM3}
\end{figure*}

\subsection{Bloch Sphere Representation of State Dynamics During QPM Sequences}
Examination of the Bloch sphere evolution of the state-vectors for individual atom pairs provides additional insight into the origin of the $1/N$ scaling of the generalized Rabi frequency as well as the mechanism underlying active dephasing suppression with QPM (see Figs. \ref{QPM1}-\ref{QPM3}). Consider several atom pairs, initially in state $pp$ at a {\it constant} detuning $E$ from resonance. At $t$=0, each experiences a different F{\"o}rster interaction strength $V(R,\Theta)$ with the same {\it constant} detuning $E \gg V(R,\Theta)$. The interaction causes the state-vectors for each atom pair to move away from $\hat{z}$ (at a polar angle $\theta$=0), precessing at similar Rabi frequencies $\Gamma(R,\Theta) \simeq E$, around axes that are tipped at small polar angles $\chi(R,\Theta) \simeq V/ E$ relative to $\hat{z}$, with azimuthal angles $\beta = \pi$. As shown in Figs. \ref{QPM1}-\ref{QPM3}, the state-vectors for the individual atom pairs trace out circles on the surface of the Bloch sphere, with the $ss'$ transition amplitudes $A_{ss'}=e^{i \varphi}\sin{\theta/2}$ defined by their respective angular coordinates $(\theta, \varphi)$. During each Rabi period, the values of $\theta$ for the different state-vectors oscillate between 0 and 2$\chi$. Due to the differing values of $\chi$ and $\Gamma$ for the different pairs, the state-vectors become angularly dispersed over a dephasing time, $\tau$. 

In a QPM sequence with $N=2$ time zones, $E \rightarrow -E$ at $t = T/2$, and the precession axes for the state-vectors jump to a new azimuthal angle, $\beta = 0$, and the precession direction reverses. The paths traced by each of the state-vectors now encircle $\hat{z}$, with radii that depend on their polar angles $\theta$ at the instant of the detuning jump. As a specific example, consider the situation where $T \simeq \pi / E$ (illustrated in Fig. \ref{QPM2} for $N$ = 4). In this case, all state-vectors complete approximately half of one precession in the first zone, such that $|A_{ss'}|$ is at (or very near) maximum with $(\theta, \varphi) \simeq (2\chi, \pi)$ at the end of the first zone ($t=T/2$). After the detuning reversal, the state-vectors rotate in the opposite direction, with cone angles 3$\chi$ relative to their {\it new} precession axes, arriving at new maximum tilt angles $(\theta, \varphi) \simeq (4\chi, 0)$ at (or very near) the end of the second time zone ($t=T$). Thus, for $N$=2 zones, the maximum transition amplitude and the time required to reach that maximum are twice as large as with constant detuning, consistent with $\Gamma \rightarrow \Gamma / 2$. Generalizing to $N$ zones (assuming $2 N \chi < 1$), $\Gamma \rightarrow \Gamma / N$ in agreement with the analytical prediction. 

Two principal factors are responsible for the suppression of dephasing through QPM. First, the variation in the Rabi frequency is significantly smaller for $E \gg E_0$, resulting in a substantial reduction in the ensemble phase variation $\Delta \phi$ over any time interval. Second, since the phase evolution is reversed in successive time zones (similar to a spin echo \cite{Hahn_echo}), $\Delta \phi$ does not accrue over the total interaction time $T$. The effect is distinct from an echo, however, because the coupling is present throughout the system evolution and $\varphi$ advances at a non-constant rate within each zone. It is also distinct from pulsed or continuous dynamical decoupling schemes in that it actively reduces dephasing due to inhomogeneities in a desirable interaction, rather than suppressing unwanted couplings to an external bath \cite{DD1,DD2,DD3,DD4,Minns_pulsed,Minns_cont,Walsworth}. Of particular importance, for each atom pair, $\varphi$ changes by $\pi$ during one Rabi cycle in zone 1 (where the path traced by the state-vector on the Bloch sphere does not enclose $\hat{z}$), as compared to 2 $\pi$ in all other zones. Therefore, the contribution to $\Delta \phi$ from zone 1 is over-corrected (by a factor of 2) by the reversed phase evolution in zone 2. So, at the end of zone 2, $\Delta \phi$ has the same magnitude, but the opposite sign as compared to that after zone 1. The same overcompensation occurs in all subsequent zones, so that at the end of the QPM sequence at $t=T$, the magnitude of $\Delta \phi$ is equal to that at the completion of zone 1, equal to that acquired during the initial interval $T/N$. Accordingly, if the ensemble dephases in a time $T = \tau$ while interacting at constant detuning then, with a QPM sequence, it will not dephase until the time spent in zone 1 is equal to $\tau$, i.e. until the total interaction time is $T= N \tau$. Stated more directly, QPM extends the dephasing time by a factor of $N$. The greater the detuning and number of zones, the greater the extension of the dephasing time. This scaling holds even for large population transfers.

Inspection of the Bloch sphere evolution of the state vectors for atom pairs with different coupling strength makes clear that the restoration of the visibility of Rabi oscillations for an ensemble is truly due to a suppression of {\it dephasing}. During any time zone, the magnitude of the probability amplitudes $|A_{ss'}|$ (as determined by $\theta$) for different atom pairs can differ significantly, but the phase angles $\varphi$ for the quantum states are nearly identical. The larger the number of zones, the smaller the time $T/N$ spent in the first zone, leading to a proportional reduction in the ensemble variation in $\varphi$.

\subsection{QPM Sequences for Suppressing Dephasing in Ordered Samples}

Quasi-phase-matching sequences can also be very effective for suppressing dephasing driven by resonant couplings in more ordered systems. One example might involve DD-interactions between atoms held at different sites in an optical lattice or tweezer array for which the internuclear separations are nominally the same, but not perfectly identical \cite{QC1,QC2,QC3}.
 
\begin{figure}
\centering
\resizebox{0.4\textwidth}{!}{\includegraphics{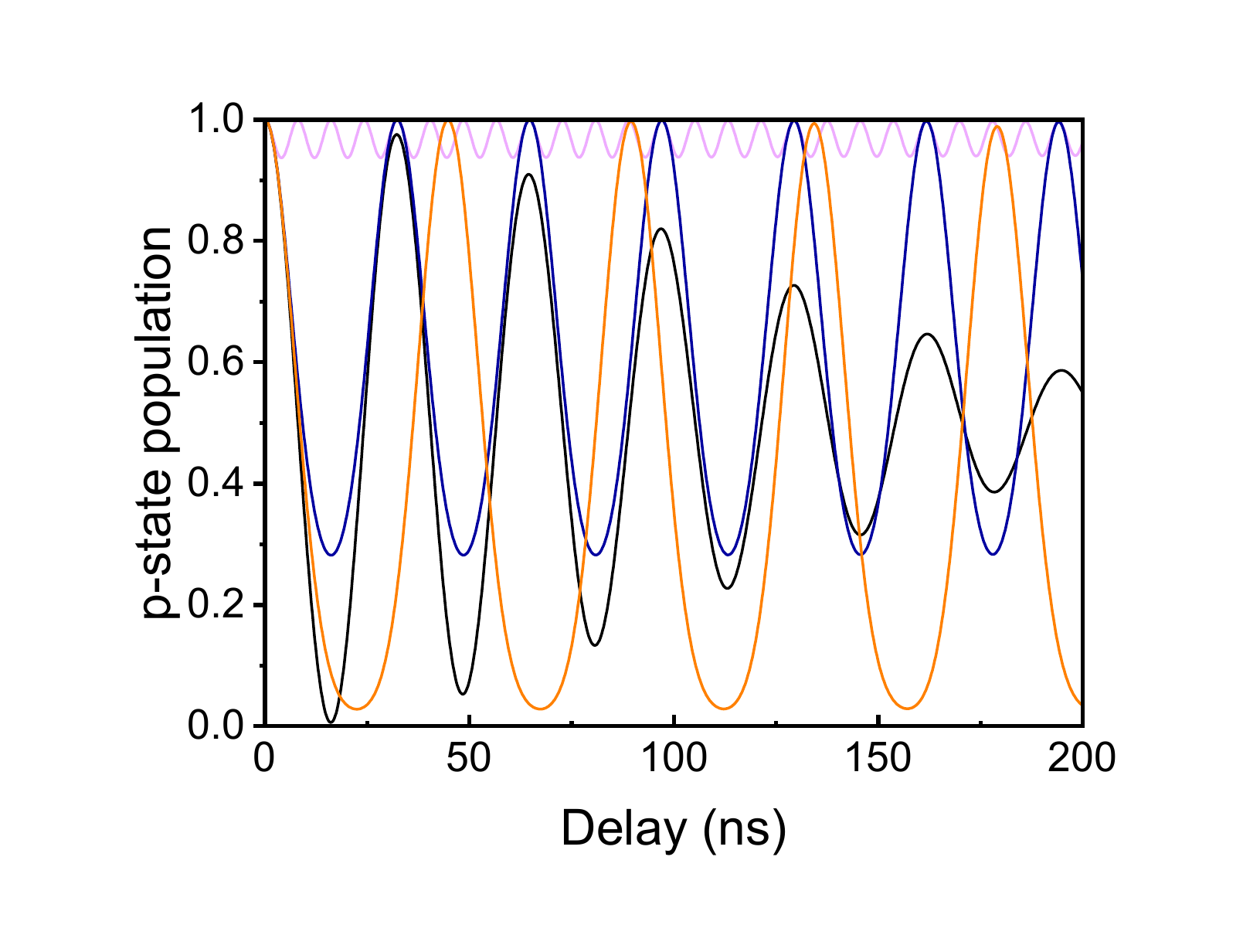}}
\caption{Simulations of the time-dependent $p$ state population (normalized to total population) for an ensemble of 10,000 isolated DD-coupled atom pairs oriented with their internuclear axes aligned along the external electric field ($\Theta = 0$), and an average internuclear distance $R$ = 3$\mu$m with a standard deviation of 50nm (Gaussian distribution). The different curves show the population transfer resulting from DD-interactions at constant detuning or pulsed QPM control sequences. The black line shows the damped Rabi oscillations that result when the atoms are tuned to the F{\"o}rster-resonance. The magenta line shows the small amplitude, high frequency Rabi oscillations predicted for a large constant detuning $E= 2\sqrt{N^2-1}V_{avg}$. The blue line shows the evolution when the atoms are subjected to a QPM sequence with $N$=4 time zones and a detuning $E= 2\sqrt{N^2-1}V_{avg}$, chosen to give the same Rabi frequency as in the resonant case. The orange curve shows the Rabi oscillations obtained using a QPM sequence with a somewhat smaller detuning, $E= 1.4\sqrt{N^2-1}V_{avg}$, allowing for higher contrast population transfer modulations ($>$97$\%$), but with a moderate reduction in the effective Rabi frequency.}
\label{ordered}
\end{figure}

Fig. \ref{ordered} shows the time-dependent $pp$ population calculated for an ensemble of 10,000 isolated DD-coupled Rb atom pairs, initially excited to the same $pp$ states relevant to our experiments, but with a Gaussian distribution of internuclear separations $R=3\pm$0.05$\mu$m and internuclear axes aligned to the external electric field axis ($\Theta = 0$). A configuration similar to this might be experimentally realized using an array of optical tweezers.

The damped Rabi oscillations seen in the black curve in Fig. \ref{ordered} reflect the significant dephasing that results from on-resonance excitation, despite the perfect pair alignment and narrow distribution of internuclear separations. The magenta curve illustrates that the dephasing can be effectively eliminated through the use of off-resonant interactions, albeit with a small maximum population transfer, if the atoms are held at a constant detuning. The blue and orange lines show that large amplitude Rabi oscillations can be restored, despite the large detunings, using QPM sequences. Further, the frequency of those oscillations can be chosen by proper selection of the detuning for a given number of QPM zones. The 97$\%$ contrast of the oscillations in the orange curve can be further improved by increasing the number of zones and/or reducing the detuning. It is worth noting, however, that the higher contrast oscillations typically deviate from perfect sinusoids (see orange curve in Fig. \ref{ordered}).  

\subsection{Simulations Including Beyond Nearest Neighbor Interactions Using 4-Atom Groups}
Although the resonant coupling between nearest neighbor atoms in an ensemble can suppress interactions with more distant atoms, that suppression is not complete and it weakens for larger detunings from resonance \cite{Robicheaux,Raithel,cusp,He}. In particular, non-tunable exchange or ``excitation hopping" interactions of the form $ps \leftrightarrow sp$, $ps' \leftrightarrow s'p$, as described through Eqs. \ref{DD_lab} and \ref{DD_full_C}, can have a non-negligible effect on the dynamics. We model these beyond nearest neighbor effects using groups of four atoms, coupled pairwise through the matrix elements given by Eqs. \ref{V++}-\ref{V-+} for the resonant processes, and through analogous expressions (involving the square of one of the two radial matrix elements $r_{ps}$ or $r_{ps'}$ rather than the product of the two) for the exchange processes. For the 4-atom groups, the problem does not reduce to one that can be solved in closed form, so the quantum dynamics within each 4-atom group are computed numerically. 

To define each four atom group, 99 atoms, each described by Cartesian coordinates $(x_i,y_i,z_i)$ are first randomly placed in a cube, the length of each edge measuring $\sqrt[3]{\frac{100}{\rho}}$, assuming a uniform distribution along each of the three spatial coordinates. We then place an atom at the center of the cube. This atom, along with its three nearest neighbors define the group of four. The relative position vectors between each pair $\vec{R_{ab}}=(R_{ab},\Theta_{ab},\Phi_{ab})$ are derived from their respective Cartesian coordinates and used to calculate the relevant matrix elements. At $t=0$, each of the four atoms are randomly assigned to one of the two possible $32p_{3/2},\abs{m_j}$=3/2 states. The probability amplitudes in all relevant 4-atom states are computed by integrating the time-dependent Schr{\"{o}}dinger equation, either for constant DD-interactions at a specific detuning $E$, or for a given QPM sequence. After an interaction time $T$, the total probabilities for finding atoms in the 32$p$, 32$s$ and 33$s$ states are recorded. Those probabilities are summed for 5000-10000 4-atom groups and then normalized to unity for comparison with the data (see Figs. 2 and 3 in the text of the primary paper). 

\bibliography{supplement}